# What Exactly is a Deepfake?

Yizhi Liu[1], Balaji Padmanabhan[1], Siva Viswanathan[1]

[1]Department of Operations & Information Technologies, University of Maryland, College Park; College Park, MD 20742, USA

## Abstract

Deepfake technologies are predominantly associated with deception, misinformation, and identity fraud, raising significant societal concerns. While these concerns are warranted, they may obscure an insight: deepfakes represent sophisticated capabilities for sensory manipulation that could alter human perceptions, thereby, perhaps counterintuitively, enabling some beneficial applications across various areas, such as healthcare and education. However, realizing this potential requires first understanding how the technology is conceptualized across different fields. This paper presents a comprehensive analysis of 826 peer-reviewed publications from 2017 to 2025, systematically examining how deepfakes are defined and understood in the literature. Using advanced large language models for content analysis, we extract and categorize deepfake conceptualizations along three key dimensions: Identity Source (the relationship between original and generated content), Intent (deceptive versus non-deceptive purposes), and Manipulation Granularity (holistic versus targeted modifications). Our analysis reveals substantial heterogeneity in deepfake conceptualizations that challenges simplified public narratives. In particular, we identify a subset of papers discussing the potential of deepfake approaches for non-deceptive intent, which has substantial yet underexplored potential for broader beneficial applications. Additionally, the temporal analysis reveals an evolution from predominantly threat-focused conceptualizations in early years (2017-2019) toward some recognition of beneficial applications in recent publications (2022-2025). This comprehensive analysis provides an empirical foundation for developing beneficial research and nuanced governance frameworks that can distinguish between applications that deserve prohibition and those that merit support and development. It suggests that, with appropriate safeguards, deepfake technology's realism can even serve important social purposes beyond its controversial origins.

## 1. Introduction

Throughout human history, technologies initially perceived as threatening often have the potential to be transformed into instruments of social benefit. Distributed computing networks, once primarily associated with malicious botnets conducting cyberattacks, now power collaborative scientific research platforms that have made crucial contributions to COVID-19 vaccine development and cancer research through projects like Folding@home (Beberg et al. 2009). Variolation, the controversial historical practice of deliberately infecting individuals with small amounts of smallpox virus to induce immunity, eventually led to the development of modern vaccination, which has saved millions of lives (Riedel 2005). Nuclear technology, despite its origins in weapons of mass destruction, has evolved to provide clean energy for millions and enable revolutionary medical treatments including radiation therapy and medical imaging (Brook et al. 2014). These historical transformations illustrate a recurring pattern: technologies with significant potential for harm often possess equally significant potential for good, with ultimate outcomes determined by how society chooses to develop, regulate, and apply them (Winner 2017).

Deepfake technology presents a contemporary manifestation of this pattern. The term "deepfake" emerged in November 2017 when a Reddit user published non-consensual, fabricated intimate content using deep learning-based face-swapping techniques, immediately establishing associations with deception, exploitation, and harm [1]. This origin has profoundly shaped subsequent discourse, with academic research, media coverage, and policy discussions focusing overwhelmingly on the technology's threats (Ajder et al. 2019). Current literature emphasizes risks including political misinformation campaigns (Chesney and Citron 2019), non-consensual intimate content (Agarwal et al. 2019), identity theft and fraud (Dolhansky et al. 2020), and the broader erosion of trust in sensory evidence (Vaccari and Chadwick 2020). Research efforts concentrate primarily on detection algorithms and defensive technologies (Li et al. 2020, Tolosana et al. 2020), while emerging regulatory frameworks seek to criminalize or severely restrict deepfake creation and distribution (Kugler and Pace 2021).

This threat-focused narrative, while addressing legitimate and critical concerns, may not fully capture the technology's fundamental capabilities and potential. At its core, deepfake technology represents a sophisticated set of techniques for manipulating sensory data (e.g., visual, audio, and multimodal content) with unprecedented realism and precision, which can alter human perceptions and elicit natural responses (Goodfellow et al. 2020, Karras et al. 2020). Such capabilities could enable transformative applications across multiple domains. For example, in medical research, deepfakes could augment training data that addresses privacy concerns while enabling algorithm training (Thambawita et al. 2021). In social science, controlled manipulation of demographic attributes in experimental stimuli could enable more rigorous studies of critical outcomes, such as bias and discrimination (Liu et al. 2025, Eberl et al. 2022). In education, historical figures could be brought to life through realistic recreations, enhancing engagement and learning outcomes (Kietzmann et al. 2020). In accessibility applications, content could be modified to better serve individuals with sensory or cognitive impairments (Giri and Brady 2023). These potential benefits remain underexplored and largely theoretical, however, because the current discourse lacks a systematic understanding of how deepfake is conceptualized and applied across different fields.

This fundamental gap in understanding motivates our research. Without comprehensive knowledge of how deepfakes are defined, categorized, and understood across academic disciplines, we cannot effectively distinguish between applications that deserve prohibition and those that merit support and development (Paris and Donovan 2019). The absence of conceptual clarity impedes both effective governance and beneficial innovation, leaving society poorly equipped to manage this powerful technology. Our research addresses this gap through two primary questions: First, *how does academic literature conceptualize deepfake technology?* Second, *what patterns in these conceptualizations might inform research and governance development?*

## 2. Background and Related Work

---

[1] Source: https://www.reddit.com/r/deepfakes/. The subreddit was created in Nov 2017 and banned in Feb 2018 after being reported by BBC, also reported by VICE.

The technical landscape of deepfake technology can be organized into three primary manipulation paradigms, each with distinct capabilities and limitations. *Face Editing* represents the most mature deepfake domain. StyleGAN3 (Karras et al. 2021) addressed the texture sticking artifacts of earlier versions while maintaining superior image quality, though it still requires substantial computational resources and struggles with unseen samples. Thus, Generative Adversarial Networks (GAN) inversion has become crucial for real image editing (Goodfellow et al. 2014). Methods like PTI (Roich et al. 2022) achieve high-quality reconstruction while maintaining editability, and e4e encoders (Tov et al. 2021) offer real-time inversion. While early inversion methods faced an inherent trade-off between reconstruction accuracy and editing flexibility (Abdal et al. 2019), recent approaches like StyleFeatureEditor (Bobkov et al. 2024) and HyperStyle (Alaluf et al. 2022) achieve an excellent balance, enabling high-fidelity reconstruction while preserving editability through regularization and optimization.

*Face Swapping* requires different architectural approaches. Modern methods like SimSwap (Chen et al. 2020) and FaceShifter (Li et al. 2020) use specialized identity encoders to preserve target attributes while transferring source identity, though they often struggle with extreme pose variations and occlusions. InfoSwap (Gao et al. 2021) improves identity preservation through information bottleneck principles, while MegaFS (Zhu et al. 2021) achieves one-shot face swapping with superior attribute preservation. Recent transformer-based approaches (Cui et al. 2023) achieve superior temporal consistency for video face swapping but at significant computational cost, limiting real-time applications. Unlike pure synthesis, these methods must balance identity preservation with natural integration. This challenge often results in subtle artifacts around boundaries or inconsistent lighting that trained observers can easily detect.

*3D-Aware and Multi-modal* advances represent emerging frontiers in deepfake methods. EG3D (Chan et al. 2022) combines GANs with neural rendering for 3D-aware face generation, enabling consistent multi-view synthesis but requiring careful camera parameter calibration. For audio-visual deepfakes, methods like Wav2Lip (Prajwal et al. 2020) achieve convincing lip synchronization, though maintaining accuracy across different languages and speaking styles remains challenging. While diffusion models like Stable Diffusion (Rombach et al. 2022) have revolutionized text-to-image generation with superior diversity and training stability, they excel primarily at creative and stylized content rather than the photorealistic human faces required for convincing deepfakes, although recent work is beginning to bridge this gap[2]. The field is also seeing the emergence of unified frameworks: vision-language models like CLIP (Radford et al. 2021) enable semantic editing through text prompts, while recent studies attempt to handle face editing with foundation models (Lu et al. 2024). However, these generalist approaches typically sacrifice performance compared to specialized methods optimized for specific tasks.

Previous surveys of deepfake technology have primarily focused on technical aspects or specific applications without systematically examining how the technology is conceptualized across disciplines. Mirsky and Lee (2021) provide a comprehensive technical survey covering creation and detection methods, categorizing approaches by architecture and training methodology. Their work offers valuable technical insights but does not address conceptual frameworks or application contexts beyond security concerns. Tolosana et al. (2020) focus specifically on facial manipulation techniques and countermeasures, providing detailed technical comparisons of detection algorithms but limited discussion of beneficial applications or conceptual diversity. Rana et al. (2022) examine deepfake detection methods through a security lens, emphasizing adversarial robustness and forensic techniques while treating all deepfakes as inherently threatening.

---

[2] Source: https://aistudio.google.com/models/gemini-2-5-flash-image, accessed on Sep 25, 2025.

The relationship between deepfake studies and broader AI-generated content research reveals conceptual confusion in the field. Truong (2024) reviews current law and policies, occasionally referencing deepfakes as a form of AI hallucination. However, hallucination typically refers to unintended factual errors or inconsistencies in AI outputs, while deepfakes involve intentional manipulation of sensory content (Ji et al. 2023). The fundamental difference between unintentional error and intentional manipulation makes this comparison conceptually flawed, yet it appears repeatedly in literature, highlighting the need for clearer conceptual frameworks. Studies of synthetic media more broadly (Fallis 2021, Paris and Donovan 2019) provide ethical frameworks for understanding manipulated content but lack systematic empirical analysis of how these technologies are actually conceptualized in practice.

Our work differs fundamentally from these prior efforts by providing the first systematic empirical analysis of how deepfake technology is conceptualized across academic literature. Rather than focusing on technical methods, detection algorithms, or philosophical implications, we map the actual definitions, categorizations, and understandings present in scholarly discourse across disciplines. This approach reveals not just what deepfakes can do technically or what ethical challenges they pose theoretically, but how different communities understand what they are and what they should be used for, providing crucial knowledge for developing appropriate governance frameworks and identifying beneficial applications.

## 3. Methodology

### 3.1 Data Collection and Filtering

To comprehensively understand how deepfake technology is conceptualized in academic literature, we designed a systematic review process spanning from November 2017 (when the term "deepfake" first appeared) through May 2025. Our methodology balances comprehensiveness with feasibility through multiple stages of filtering and validation, ensuring both broad coverage and analytical rigor. We initiated our search using Google Scholar due to its comprehensive coverage across disciplines. The initial query combining "deepfake" and "artificial intelligence" as search terms yielded 17,200 results, reflecting the technology's broad impact across academic fields. Recognizing the infeasibility of manually analyzing this entire corpus while maintaining quality standards, we developed a systematic sampling strategy. Using Publish or Perish 8.0 software[3], we extracted bibliographic information for the top 1,000 papers ranked by Google Scholar's relevance algorithm. The software enabled systematic extraction of metadata including titles, authors, publication years, venues, citation counts, and source URLs.

Our inclusion criteria encompassed peer-reviewed journal articles and conference proceedings, with a focus on English-language publications to ensure consistent analysis. We further applied several exclusion criteria to maintain analysis quality and consistency: papers with zero citations published before 2025 (as a quality threshold while allowing recent papers), retracted publications, book chapters and technical reports to focus on peer-reviewed content, and duplicate entries. This filtering process yielded 826 unique papers with accessible full text. Document acquisition proceeded through multiple channels. We imported the filtered bibliographic data into Zotero reference management software for systematic organization. Through institutional subscriptions and automated retrieval tools, we successfully downloaded 677 papers (82%). The remaining 149 papers required manual acquisition through ResearchGate, publisher platforms, or preprint servers (e.g., arXiv and SSRN).

---

[3] Source: https://harzing.com/blog/2021/10/publish-or-perish-version-8, accessed on Sep 25, 2025.

## 3.2 Framework Development

To systematically categorize how deepfakes are conceptualized in the literature, we developed a three-dimensional framework through iterative discussions with GPT-5 and validation by domain experts, as shown in Table 1. The *Identity Source* dimension captures the relationship between source and generated content, distinguishing between: Original-to-Target Transfer (A→B) where one identity completely replaces another (e.g., face-swapping); Original Modification (A→A') where attributes are altered while preserving core identity (e.g., age progression, expression change); and Complete Synthesis (∅→B) involving creation of entirely artificial identities without real-world referents (e.g., synthetic actors).

**Table 1. Three dimensions of deepfake definitions**

| Identity Source | Intent | Manipulation Granularity |
|---|---|---|
| Original-to-Target Transfer (A→B) | Deceptive | Targeted |
| Original Modification (A→A') | Non-deceptive | Holistic |
| Complete Synthesis (∅→B) | - | - |

The *Intent* dimension distinguishes the purpose behind deepfake creation: Deceptive applications intended to mislead people about authenticity (e.g., misinformation, fraud); and Non-deceptive uses created for legitimate purposes with transparency about artificial nature (e.g., entertainment, research, education). The *Manipulation Granularity* dimension characterizes the scope of alterations: Holistic transformations involving comprehensive changes affecting entire identity or scene (e.g., complete identity replacement); and Targeted modifications involving selective alteration of specific attributes or features (e.g., changing only eye color or facial expression). This framework, grounded in both technical capabilities and application contexts, provides a structured approach for analyzing the diverse conceptualizations present in academic literature.

## 3.3 Definition Extraction and Categorization

For systematic content analysis, we employed DeepSeek-R1[4], a state-of-the-art large language model with explicit reasoning capabilities and outputs that enable verification of extraction and categorization results. We developed a comprehensive prompt (see Appendix Figure A1), instructing the model to extract from each paper: the paper's deepfake definition (explicit or inferred), primary applications mentioned, categorization of deepfake definitions for each of the three dimensions (i.e., Identity Source, Intent, Manipulation Granularity), and reasoning explanations justifying each categorization. These reasoning traces proved crucial for validation, allowing us to verify that categorizations aligned with actual paper content rather than model hallucinations. The extraction process produced structured JSON outputs for ease of analysis.

---

[4] To address copyright and privacy concerns, we used a local version of DeepSeek-R1 via Ollama: https://ollama.com/library/deepseek-r1, accessed on Oct 5, 2025.

We preserved and analyzed the model's reasoning traces for internal consistency, flagging 120 papers where reasoning suggested one categorization but the output indicated another. These conflicts underwent manual review, with corrections based on direct textual evidence from the source papers. Papers could be assigned to multiple categories within each dimension if they discussed various types of deepfakes (e.g., a paper might discuss both deceptive and non-deceptive applications). The combination of systematic framework development, automated extraction with reasoning traces, and rigorous validation ensures that our analysis accurately captures the conceptual landscape of deepfake technology across academic literature.

## 4. Results

### 4.1 Overall Distribution of Conceptualizations

Our analysis of 826 papers yielded 659 containing explicit or implicit deepfake definitions, while 167 papers discussed the technology without providing clear conceptualizations. The 659 papers with definitions form the basis of our subsequent analysis. As illustrated in Figure 1, the distribution across our three-dimensional framework reveals heterogeneity in how deepfake technology is conceptualized, though with stronger concentration in threat-focused categories. Analysis of the Identity Source dimension reveals overwhelming emphasis on Original-to-Target Transfer applications, appearing in 628 papers (95.3% of papers with definitions). This predominance reflects the technology's origins in face-swapping applications popularized by the 2017 Reddit incident. Original Modification appears in 60 papers (9.1%), representing limited recognition of applications where identity is preserved but attributes are altered. Complete Synthesis appears in 39 papers (5.9%), where entirely fictional characters are created.

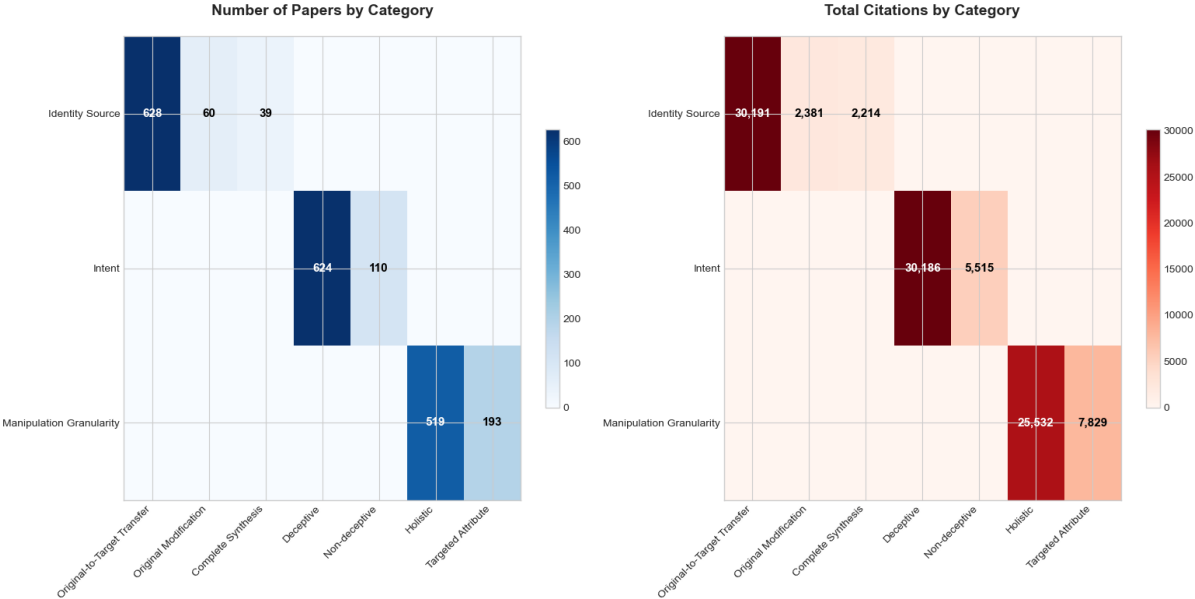

**Figure 1. Distribution of Papers and Citations by Dimensional Categories. Note:** The total number of papers are over 659 since one paper may contain multiple definitions.

The Intent dimension reveals the dominance of deceptive conceptualizations. Deceptive applications appear in 624 papers (94.7%), while Non-deceptive uses are discussed in only 110 papers (16.7%). This stark imbalance reflects the field's focus on threats, with papers studying deepfakes primarily through a security and detection lens. The overlap (papers discussing both intents) suggests some recognition that the same technical capabilities can serve different purposes, though this remains a relatively minor perspective. The Manipulation Granularity dimension shows a strong preference for Holistic transformations in 519 papers (78.8%), reflecting the predominant understanding of deepfakes as complete identity or scene replacements. Targeted modifications appear in 193 papers (29.3%), suggesting limited but present appreciation for more nuanced manipulation capabilities. The targeted modification papers span diverse applications from medical image augmentation preserving diagnostic features, to experimental psychology studies manipulating specific facial attributes.

## 4.2 Cross-Dimensional Analysis

As shown in Table 2, cross-tabulation of the three dimensions reveals eight primary conceptual categories of deepfake, with papers potentially appearing in multiple categories if they discuss various types. The most populated combination involves Original-to-Target Transfer with Deceptive Intent and Holistic Manipulation (470 papers, 71.3%), representing the typical malicious face-swap that dominates both public perception and academic discourse. This category includes papers on non-consensual intimate content (Agarwal et al. 2019), political misinformation (Chesney and Citron 2019), and identity theft (Dolhansky et al. 2020).

**Table 2. Deepfake Definitions by Dimensions. Note:** The total number of papers are over 659 since one paper could have multiple definitions. A complete list of the 659 papers and their deepfake definitions are provided in Table A1.

| Identity | Intent | Gran. | Papers (Cites) | Application(s) | LLM Summarized Definition |
|---|---|---|---|---|---|
| A→B | Deceptive | Holistic | 470 (23,156) | Misinformation & Identity Theft (Nguyen et al. 2019) | AI-driven technology designed for deception, where complete identities are transferred from an original source to a target using holistic manipulation, aiming to mislead individuals with malicious intent. |
| A→B | Deceptive | Targeted | 145 (6,982) | Targeted Manipulation Attacks (Prezja et al. 2022) | AI-generated content where specific targeted attributes from one source are manipulated to create deceptive media, focusing on precise alterations of particular characteristics like voice tone, facial features, or expressions. |
| A→B | Non-Deceptive | Holistic | 28 (1,247) | Research & Entertainment (Eberl et al. 2022) | Technology that transfers complete identities from original sources to targets with non-deceptive intent, used for legitimate purposes including academic research, entertainment, and educational applications. |
| A→A' | Deceptive | Targeted | 42 (2,068) | Attribute-Level Spoofing (Khochare et al. 2021) | Technology that generates synthetic content by modifying specific attributes of original material, designed to deceive by targeting particular characteristics like biological features. |
| A→A' | Non-Deceptive | Holistic | 8 (412) | Research Enhancement (Thambawita et al. 2021) | Technology that modifies original data to generate synthetic content without deceptive intent, focusing on creating realistic alternatives for medical, educational, or research areas. |
| A→A' | Non-Deceptive | Targeted | 6 (285) | Content Enhancement | Technology that creates synthetic content by modifying specific attributes of original data for |

| | | | | (Herring et al. 2022) | legitimate purposes like beauty filters and entertainment. |
|---|---|---|---|---|---|
| ∅→B | Deceptive | Holistic | 31 (1,759) | Synthetic Identity Creation (Xie et al. 2024) | Technology that creates entirely synthetic identities from scratch without relying on real-world data, designed to deceive viewers into believing the fabricated content represents authentic individuals or events. |
| ∅→B | Non-Deceptive | Holistic | 8 (455) | Legitimate Synthetic Content (Kaate et al. 2024a) | Technology that creates entirely new synthetic content from scratch with transparent, non-deceptive intent, designed for legitimate entertainment, research, or creative applications. |

Beyond this dominant category, two combinations merit particular attention. Studies employing Original-to-Target Transfer with Non-deceptive Intent and Holistic Manipulation (28 papers, 4.2%) represent the closest non-malicious parallel to the dominant deceptive category, using similar technical approaches but for legitimate purposes. These include research and entertainment applications (Eberl et al. 2022) and educational uses. The existence of this category, though small, demonstrates that complete identity transfer technology can serve beneficial purposes when implemented with transparency and appropriate consent.

In addition, the combination of Original Modification (A→A'), Non-deceptive Intent, and Targeted Manipulation contains only 6 papers (0.9%), representing the intersection highly suited for beneficial applications. This approach, modifying specific attributes of original content rather than complete replacement, inherently involves greater controllability. The few papers in this category focus on content enhancement applications (Herring et al. 2022), such as beauty filters and entertainment purposes. Despite the theoretical potential for beneficial applications using precise attribute control while maintaining transparency, such as experimental stimuli in scientific research, this category remains underexplored. These non-deceptive applications remain a small minority compared to deceptive ones using a similar technical approach (e.g., audio spoofing).

## 4.3 Temporal Evolution

Analysis of publication dates reveals clear temporal patterns in how deepfakes are conceptualized. Based on our 659 papers with definitions, we identify three distinct phases in the evolution of deepfake conceptualizations. As illustrated in Figure 2, the period from 2018 to 2019 represents the "Threat Emergence" phase, with 23 papers published containing definitions. During this period, pure non-deceptive applications were entirely absent (0%), with all papers focusing on deceptive uses or both deceptive and non-deceptive applications. When including papers discussing both types, non-deceptive applications appeared in 17.4% of papers. Technical papers emphasized detection and forensics (Li et al. 2020), while application papers concentrated on non-consensual intimate content and political misinformation.

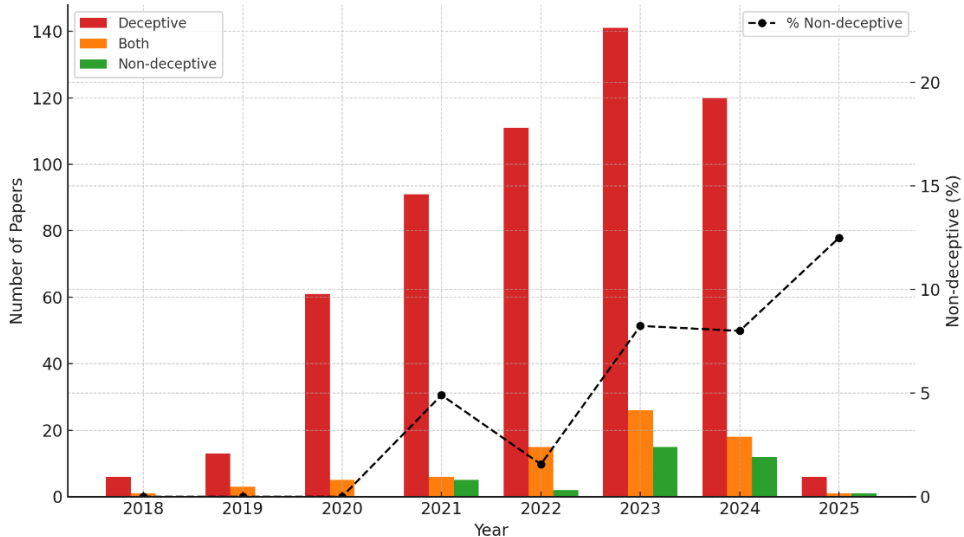

**Figure 2. Temporal Evolution of Deepfake Conceptualization**

From 2020 to 2021, the "Pandemic Adaptation" phase encompassed 168 papers. Pure non-deceptive applications emerged but remained minimal at 3.0% of papers (5 papers), while papers discussing both deceptive and non-deceptive uses contributed to a total of 9.5% mentioning non-deceptive applications. Medical applications gained some attention as researchers explored synthetic data for training diagnostic algorithms while preserving patient privacy during a period of rapid telehealth expansion (Thambawita et al. 2021). As responses to social distancing requirements, papers exploring deepfake-based virtual presence also emerged (Wu et al. 2021).

The period from 2022 to 2024 represents "Gradual Diversification," with 460 papers demonstrating modest evolution. Pure non-deceptive applications increased to 6.3% of papers (29 papers), while the total discussing non-deceptive uses (including papers addressing both) reached 19.1%. This represents real but limited growth in beneficial conceptualizations. The period sees the emergence of sophisticated applications including medical imaging research, such as Prezja et al. (2022) demonstrating deepfake-generated knee osteoarthritis X-rays for augmenting medical training datasets, and Kim et al. (2022) exploring deepfake detection specifically for medical data integrity. Legal and governance discussions evolved beyond simple prohibition, with papers like Custers (2022) examining AI applications in criminal law contexts and Lewis (2022) proposing comprehensive legal frameworks for AI regulation. Research methodology applications expanded, including Kleinlogel et al. (2023) using AI-generated standardized corpora for research purposes. These developments suggest growing recognition that deepfake technology requires nuanced governance approaches, distinguishing between harmful and beneficial applications, though security concerns continue to dominate the discourse.

## 5. Conclusion and Discussion

This systematic analysis of 826 academic papers reveals that deepfake technology is overwhelmingly conceptualized through a threat-focused lens. Through our three-dimensional framework encompassing Identity Source, Intent, and Manipulation Granularity, we identify that 94.7% of papers discuss deceptive applications while only 5.3% focus purely on non-deceptive uses. However, the temporal evolution from predominantly threat-focused conceptualizations in 2017-2019 toward increasing recognition of beneficial applications in 2022-2025 suggests that beneficial uses of deepfakes are possible, and the exposure and experience foster a more nuanced understanding.

As deepfake technology continues its rapid evolution, maintaining awareness of conceptual diversity becomes increasingly critical for researchers, policymakers, and society. The academic community's growing recognition of beneficial applications suggests that with appropriate frameworks and safeguards, deepfake technology's sophisticated sensory manipulation capabilities can serve important social purposes far beyond its controversial origins. The transformation of deepfakes from threat to tool will require continued research, thoughtful governance, and public engagement, but our analysis provides compelling evidence that this transformation is not only possible but already underway within academic discourse. The question is not whether deepfakes can be beneficial, but how quickly and effectively we can develop the frameworks necessary to realize their positive potential while managing their risks.

### 5.1 Implications for Researchers and Governance

Despite the increasing recognition of deepfakes' potential for beneficial applications, our analysis suggests isolated efforts rather than a coherent research program. For instance, while Thambawita et al. (2021) demonstrate deepfake-based electrocardiogram generation preserving diagnostic features while protecting privacy, and Eberl et al. (2022) show deepfakes enabling identification of bias through experiments, these pioneering works have not spawned substantial follow-up research. This gap may stem from multiple factors. The first might be technical barriers because most available deepfake tools are designed for entertainment or malicious purposes, requiring specific modification for research applications. Additionally, some sophisticated systems remain proprietary and inaccessible to researchers. Creating custom research-grade systems requires interdisciplinary collaborations and expertise in deep learning, computer vision, and domain-specific knowledge. The reputational concerns about association with a controversial technology create additional barriers, as researchers may worry that deepfake-related work might hinder career prospects. This requires recognizing that the same techniques can serve legitimate and beneficial needs, with careful ethical guidance and safeguards.

Our analysis finds conceptual diversity of deepfakes, which are different from the current governance frameworks that mainly advocate the prohibition of deepfakes wholesale. Proposed legislation like the European Union's AI Act[5] and various US state bills[6] treat deepfakes as a monolithic category, failing to distinguish between different types of applications with vastly different risk profiles and benefit potentials (Kugler and Pace 2021). Our three-dimensional framework offers a foundation for more sophisticated governance that recognizes this diversity.

Identity relationship governance could establish different regulatory requirements based on the Identity Source dimension. Original-to-Target Transfer applications might require explicit consent from both source and target individuals, with stringent penalties for non-consensual use. Original Modification applications could operate under lighter requirements, requiring only the subject's consent for legitimate purposes. Complete Synthesis applications might face minimal restrictions when creating entirely fictional entities. This differentiated approach would protect individual rights while enabling beneficial applications that don't involve identity substitution.

---

[5] Source: https://www.europarl.europa.eu/topics/en/article/20230601STO93804/eu-ai-act-first-regulation-on-artificial-intelligence, accessed on Sep 25, 2025.
[6] Source: https://www.ncsl.org/technology-and-communication/deceptive-audio-or-visual-media-deepfakes-2024-legislation, accessed on Sep 25, 2025.

Intent-based regulation could focus on deceptive intent and actual harm rather than the technology itself. This parallels existing legal frameworks for fraud, defamation, and misrepresentation that criminalize deceptive intent regardless of medium. Non-deceptive applications with clear disclosure could operate under transparency standards similar to those governing edited photographs in journalism or CGI in films. This approach would avoid criminalizing the technology while addressing actual harms from deception.

Manipulation granularity could inform oversight intensity, with targeted modifications for specific research or medical purposes receiving lighter regulation than holistic identity transformations. For instance, research applications modifying specific attributes could follow existing institutional review board protocols for human subject research, while entertainment applications creating holistic changes might require different disclosure standards. This granular approach would enable proportionate governance aligned with actual risks and benefits.

## 5.2 Limitations and Future Directions

Several limitations of our study warrant acknowledgment. Our focus on English-language academic literature may miss important perspectives from other linguistic communities, particularly given significant deepfake research in non-English speaking countries, where different cultural attitudes toward synthetic media might yield alternative conceptualizations. The rapid evolution of deepfake technology means our analysis captures a snapshot of a moving target. New architectures like diffusion models and neural radiance fields continue to emerge, potentially enabling applications not yet imagined. Our analysis ending in May 2025 might miss important recent developments. While our LLM-based extraction method enabled analysis of a large corpus efficiently, it may miss subtle conceptual nuances that close human reading would capture.

Future research should address these limitations through several complementary approaches. Cross-cultural analysis of non-English literature could reveal whether our findings reflect universal patterns or Anglo-centric perspectives. Beyond academia, engaging with industry developers, policymakers, and civil society organizations would provide a more comprehensive understanding of how different stakeholders conceptualize deepfake technology. Longitudinal studies could identify specific factors, such as technological advances, regulatory changes, or social events, that drive conceptual evolution. Moving from theory to practice, empirical validation of beneficial applications we identify could demonstrate actual value rather than theoretical potential. Developing open-source, purpose-built research tools would directly address the technical barriers preventing beneficial applications. Finally, systematic impact assessments comparing harmful and beneficial uses could provide the evidence base necessary for nuanced governance frameworks that promote innovation while managing risks.

# Appendix

## Table A1. Deepfake Categorization and Definitions

| Num. | Year | Author(s) | Source | Intent | Granularity | Definition | Citation |
|---|---|---|---|---|---|---|---|
| 1 | 2018 | Badale et al. | A→B | Deceptive | Targeted | Artificially generated videos that use AI to merge or superimpose existing images or videos in a realistic manner. | 28 |
| 2 | 2018 | Bates | A→B | Deceptive | Holistic | Artificially generated videos that realistically swap faces between people, making them difficult to distinguish from real content. | 9 |
| 3 | 2018 | Fletcher | A→B | Deceptive and Non-deceptive | Holistic | Artificially generated images/videos using AI to swap or alter faces for deception. | 195 |
| 4 | 2018 | Güera and Delp | A→B | Deceptive | Holistic | Artificially generated videos that involve face swaps to create manipulated content for deceptive purposes. | 1389 |
| 5 | 2018 | Hall | A→B | Deceptive | Holistic | Artificially generated videos that manipulate or swap identities to deceive viewers. | 48 |
| 6 | 2018 | Hauser and Ruef | A→B | Deceptive | Holistic | Deepfakes are artificial face-swapping technologies that replace one person's face with another while maintaining realistic expressions and surroundings. | 4 |
| 7 | 2018 | Koopman et al. | A→B | Deceptive | Targeted | Deepfakes are used to manipulate video evidence in legal contexts by altering faces to impersonate individuals. | 214 |
| 8 | 2019 | Đorđević et al. | A→B | Deceptive | Targeted and Holistic | Artificially generated videos that involve face swapping or identity replacement, analyzed using SIFT features to detect them. | 12 |
| 9 | 2019 | Albahar and Almalki | A→B | Deceptive and Non-deceptive | Targeted | Deepfakes are tools that create convincing fake images or videos through machine learning to manipulate identities and spread misleading information. | 139 |
| 10 | 2019 | Bazarkina and Pashentsev | A→B | Deceptive | Targeted | Deepfakes involve the use of AI to manipulate or replace parts of an individual's identity for deceptive purposes. | 68 |
| 11 | 2019 | Byreddy | A→B | Deceptive and Non-deceptive | Targeted | Artificially generated videos that use AI techniques to alter or impersonate identities, often for deceptive purposes. | 3 |
| 12 | 2019 | Caldera | A→B | Deceptive | Holistic | Artificially generated videos that manipulate existing content to create realistic or altered representations. | 58 |
| 13 | 2019 | Dolhansky et al. | A→B | Deceptive | Targeted | Artificially manipulated videos where facial identities are altered or swapped for deceptive purposes. | 684 |
| 14 | 2019 | Durall et al. | A→B | Deceptive | Holistic | AI-generated face images that appear realistic but are artifacts of deep learning models. | 304 |
| 15 | 2019 | Fernandes et al. | A→B | Deceptive | Holistic | Deepfakes are artificial videos where faces are replaced to create manipulated images or expressions. | 142 |
| 16 | 2019 | Hasan and Salah | A→B | Deceptive | Holistic | Artificially generated videos that use AI and deep learning to copy and enhance content, making it appear authentic. | 388 |
| 17 | 2019 | LaMonaca | A→B | Deceptive and Non-deceptive | Holistic | Artificially generated videos that can impersonate real identities or scenes, often used to challenge authentication standards. | 22 |
| 18 | 2019 | Maras and Alexandrou | A→B | Deceptive | Holistic | Deepfakes are AI-generated videos that use techniques like face swapping to create fake or altered content. | 446 |
| 19 | 2019 | Nguyen et al. | A→B | Deceptive | Holistic | Deepfakes are artificial images or videos created using deep learning algorithms that mimic real ones to the extent that humans cannot distinguish them. | 219 |
| 20 | 2019 | Pfefferkorn | A→B | Deceptive | Holistic | Deepfakes are created to impersonate individuals or events, often used for authentication purposes. | 79 |
| 21 | 2019 | Sohrawardi et al. | A→B | Deceptive | Holistic | AI-enhanced media used for deception or fake announcements. | 64 |
| 22 | 2019 | Weiss | A→B | Deceptive | Holistic | Deepfakes are generated text comments that appear human but aren't based on conventional patterns. | 38 |

| 23 | 2019 | Westerlund | A→B | Deceptive | Holistic | Deepfakes are AI-generated videos that use face swapping to create realistic manipulations of identities. | 1309 |
|----|------|-----------|-----|-----------|----------|------|------|
| 24 | 2020 | Đorđević et al. | A→B | Deceptive | Holistic | Artificially generated videos that replace or modify faces and identities using AI and deep learning techniques. | 4 |
| 25 | 2020 | Amerini and Caldelli | A→B | Deceptive | Holistic | Synthetic videos with altered facial expressions or identities that are detected using machine learning techniques focusing on prediction errors in frames. | 73 |
| 26 | 2020 | Buo | A→B | Deceptive | Holistic | Deepfakes are AI-based technologies that manipulate images, sounds, or videos to represent non-existent events or identities. | 32 |
| 27 | 2020 | Chan et al. | A→B | Deceptive | Targeted | Deepfakes are artificially generated digital content designed to deceive by altering or swapping identities. | 59 |
| 28 | 2020 | Chang et al. | A→B | Deceptive | Holistic | Artificially generated face images that are tampered with to appear real, often using deep learning techniques to detect or mimic such alterations. | 97 |
| 29 | 2020 | Charitidis et al. | A→B | Deceptive | Holistic | Deepfakes involve AI-generated manipulated media to deceive or impersonate individuals. | 31 |
| 30 | 2020a | Chintha et al. | A→B | Deceptive | Holistic | Artificially generated videos that manipulate faces to create deceptive content. | 43 |
| 31 | 2020b | Chintha et al. | A→B | Deceptive | Holistic | Artificially generated audiovisual renderings that realistically mimic real content to deceive or manipulate public perception. | 217 |
| 32 | 2020 | Dolhansky et al. | A→B | Deceptive | Holistic | Artificially generated face-swapped videos used for deepfake detection research. | 1024 |
| 33 | 2020 | Dunn | A→B | Deceptive | Holistic | Deepfakes involve the use of technology to replicate aspects of people's identities, such as faces and voices. | 21 |
| 34 | 2020 | Farish | A→B | Deceptive | Holistic | Deepfakes are artificial methods used to swap or replace identities in images or videos, often for deceptive purposes. | 30 |
| 35 | 2020 | Feng et al. | A→B | Deceptive | Holistic | Deepfakes are videos where faces have been tampered with or replaced using advanced AI techniques, often to deceive viewers. | 8 |
| 36 | 2020 | Frank et al. | A→B | Deceptive | Targeted | Deepfakes are artificial images generated by Generative Adversarial Networks (GANs) that exhibit artifacts in the frequency domain, which can be used to identify them. | 704 |
| 37 | 2020 | Gandhi and Jain | A→B | Deceptive | Holistic | Deepfakes created using adversarial perturbations to deceive deepfake detectors by altering images in ways that bypass detection methods. | 169 |
| 38 | 2020 | Geddes | A→B | Deceptive | Targeted and Holistic | Deepfakes are artificial images or videos that use realistic face swapping to impersonate individuals. | 27 |
| 39 | 2020 | Gerstner | A→B | Deceptive | Holistic | Artificially generated videos that swap faces between individuals using AI technology. | 46 |
| 40 | 2020 | Gino | A→B | Deceptive | Targeted | Synthetic media generated using AI to deceive by altering identities or expressions. | 3 |
| 41 | 2020 | Gong et al. | A→B | Deceptive | Holistic | Artificially generated images or videos that involve face swapping or identity manipulation to deceive. | 37 |
| 42 | 2020 | Gosse and Burkell | A→B | Deceptive and Non-deceptive | Holistic | Deepfakes are created by altering faces in videos to create false identities or images. | 117 |
| 43 | 2020 | Graber-Mitchell | A→B | Deceptive | Targeted | Artificially generated videos depicting deceptive or harmful scenarios, such as non-consensual intimate interactions. | 7 |
| 44 | 2020a | Guarnera et al. | A→B | Deceptive | Holistic | Deepfakes are synthetic images or videos where faces are replaced using deep learning algorithms to mimic real identities. | 73 |
| 45 | 2020b | Guarnera et al. | A→B | Deceptive | Holistic | Deepfakes are artificial images created using deep learning techniques that manipulate face identities through transformations like gender swaps and aging. | 144 |
| 46 | 2020 | Gupta et al. | A→B | Deceptive | Holistic | Synthetic videos designed to deceive viewers by altering or replacing visual elements such as faces. | 46 |
| 47 | 2020 | Hongmeng et al. | A→B | Deceptive | Targeted | Artificially generated videos that use techniques like face swapping and super-resolution reconstruction to create realistic or altered identities. | 22 |
| 48 | 2020 | Ivanov et al. | A→B | Deceptive | Targeted | AI-generated images or videos with manipulated identities, often used to create realistic fake images or videos. | 29 |
| 49 | 2020 | Jafar et al. | A→B | Deceptive | Targeted and Holistic | Deepfakes are detected using mouth features to identify fake videos by analyzing lip/mouth movement. | 120 |
| 50 | 2020 | Jaiman | A→B | Deceptive and Non-deceptive | Holistic | Artificially generated content that realistically swaps or modifies identities for deceptive purposes. | 20 |
| 51 | 2020 | Jones | ∅→B | Deceptive | Holistic | Deepfakes are AI-generated fake content that manipulate identities or attributes to deceive users. | 26 |
| 52 | 2020 | Jung et al. | A→B | Deceptive | Holistic | Deepfakes are AI-generated images or videos that mimic real people's identities through techniques like GANs, often used for deceptive purposes. | 313 |
| 53 | 2020 | Karnouskos | A→B | Deceptive and Non-deceptive | Holistic | Deepfakes are realistic digital products created using artificial intelligence for deceptive purposes, often involving the creation of fake videos with altered or swapped identities. | 276 |
| 54 | 2020 | Kaur et al. | A→B | Deceptive | Holistic | Deepfakes are video or image forgery techniques that use neural networks to superimpose existing images onto destination videos, often for spreading misinformation. | 45 |
| 55 | 2020 | Kawa and Syga | A→B | Deceptive | Holistic | Artificially generated videos that realistically swap faces between people using neural networks to impersonate individuals. | 17 |
| 56 | 2020 | Kietzmann et al. | A→B | Deceptive and Non-deceptive | Targeted | Deepfakes are AI-generated content that can deceive by altering visual or auditory media, making them harder to detect. | 583 |
| 57 | 2020 | Kirchengast | A→B | Deceptive | Holistic | AI-based technology that alters video images to replace or manipulate identities. | 81 |
| 58 | 2020 | Langer and Wyczik | A→B | Deceptive | Holistic | Deepfakes are artificial images or videos that realistically swap faces between individuals. | 3 |
| 59 | 2020 | Liang et al. | A→B | Deceptive | Holistic | DeepFakes are artificially generated videos that involve the replacement or manipulation of identities through face swapping or other means to deceive. | 10 |
| 60 | 2020 | Lomnitz et al. | A→B | Deceptive | Holistic | Deepfakes are synthetic media created using Generative Adversarial Networks (GANs) to mimic real data for nefarious purposes such as identity replacement or impersonation. | 29 |
| 61 | 2020 | Maddocks | A→B | Deceptive | Targeted | AI-generated images where faces are swapped or altered, often used for deceptive purposes. | 170 |
| 62 | 2020 | Maksutov et al. | A→B | Deceptive | Holistic | Deepfakes involve AI-generated alterations of identities, particularly through face swapping or identity replacement. | 82 |
| 63 | 2020 | Malolan et al. | ∅→B | Deceptive | Holistic | Artificially generated videos that swap faces between individuals to deceive or impersonate others. | 69 |
| 64 | 2020 | Mitra et al. | A→B | Deceptive | Holistic | AI-generated videos used to deceive or impersonate individuals. | 59 |
| 65 | 2020 | Mittal et al. | A→B | Deceptive | Holistic | Artificially generated content that manipulates audio and visual modalities to deceive through emotional cues. | 351 |
| 66 | 2020 | Muna | A→B | Deceptive | Holistic | Artificially generated media that uses deep learning to alter or swap identities. | 4 |
| 67 | 2020 | Nagel | A→B | Deceptive | Holistic | Manipulated images used to control or verify women's images in digital spaces. | 57 |
| 68 | 2020 | Nasar et al. | A→B | Deceptive | Holistic | Deepfakes are artificial digital manipulations used to create realistic videos depicting people's actions or identities that never occurred. | 25 |

| 69 | 2020 | Nazar and Bustam | A→B | Deceptive | Holistic | Deepfakes are AI-generated videos that can replace or alter identities, making them appear as if they were real. | 17 |
| 70 | 2020 | Pantserev | A→B | Deceptive | Holistic | AI-generated images or videos that manipulate identities, often for malicious purposes. | 115 |
| 71 | 2020 | Pashentsev | A→B | Deceptive | Holistic | Artificially generated content used for deceptive or harmful political purposes. | 16 |
| 72 | 2020 | Repez and Popescu | Mixed Types | Deceptive | Holistic | Deepfakes are artificial face swaps used to deceive individuals or groups in political contexts. | 7 |
| 73 | 2020 | Singh et al. | A→B | Deceptive | Holistic | Artificially generated videos that mimic real content but are forged, often used to deceive. | 57 |
| 74 | 2020 | Singh et al. | A→B | Deceptive | Holistic | Deepfakes are artificial images or videos used to deceive by swapping or modifying identities. | 29 |
| 75 | 2020 | Sohrawardi et al. | A→B | Deceptive and Non-deceptive | Holistic | AI-generated manipulated audio or video that appears real but involves deceptive identity replacement. | 18 |
| 76 | 2020 | Suratkar et al. | A→B | Deceptive | Targeted and Holistic | Artificially generated fake videos that mimic real content for harmful purposes. | 31 |
| 77 | 2020 | Temir | A→B | Deceptive | Targeted and Holistic | AI-generated realistic videos or audio where identities or content are altered. | 47 |
| 78 | 2020 | Usukhbayar and Homer | A→B | Deceptive | Holistic | Deepfakes are artificial video manipulations that use AI to create realistic representations of other people's identities, often through face swapping or attribute modification. | 8 |
| 79 | 2020 | Vaccari and Chadwick | A→B | Deceptive | Targeted | Synthetic videos used to spread disinformation, often involving face swapping or identity replacement. | 862 |
| 80 | 2020 | Whittaker et al. | A→B | Deceptive | Holistic | Deepfakes are AI-generated images or videos used to create synthetic faces that appear real. | 108 |
| 81 | 2020 | Whyte | A→B | Deceptive | Targeted and Holistic | AI-generated content that is difficult for humans and machines to distinguish from real content, often involving face swapping or identity replacement. | 101 |
| 82 | 2020 | Wu et al. | A→B | Deceptive | Holistic | Deep fake images are artificially generated images where faces are manipulated using AI techniques to swap identities or alter features. | 14 |
| 83 | 2020 | Yagnik et al. | A→B | Deceptive | Holistic | Deepfakes are artificial audiovisual content created using GANs to transpose faces onto targets, often for realistic and convincing results. | 4 |
| 84 | 2020a | Younus and Hasan | A→B | Deceptive | Holistic | Deepfakes are synthetic videos created using AI to replace or modify parts of real footage, often aiming to deceive viewers. | 16 |
| 85 | 2020b | Younus and Hasan | A→B | Deceptive | Holistic | AI-generated fake videos with manipulated faces that require additional blur to fit into the background context of source videos. | 60 |
| 86 | 2020 | Zhang et al. | A→B | Deceptive | Holistic | AI-generated fake images or videos used for various applications like entertainment. | 33 |
| 87 | 2020 | Zhao et al. | A→B | Deceptive | Holistic | Artificially generated videos that involve face swapping and tampering to deceive or impersonate individuals. | 23 |
| 88 | 2020 | Zi et al. | A→B | Deceptive | Holistic | Artificially generated videos that involve realistic face swapping to test deepfake detection systems. | 487 |
| 89 | 2020 | Zotov et al. | A→B | Deceptive | Targeted | DeepFakes are images or videos where faces are manipulated using AI techniques like GANs and CNNs to create realistic alterations. | 11 |
| 90 | 2021 | Agarwal and Singh | A→B | Deceptive | Targeted | AI-generated fake media with face swaps that are difficult to distinguish from real images. | 27 |
| 91 | 2021 | Agnihotri | A→B | Deceptive | Holistic | Deepfakes are artificial images or videos created using algorithms like Generative Adversarial Networks (GANs) to swap facial features between individuals, resulting in realistic but fake identities. | 4 |
| 92 | 2021 | Ahmed et al. | A→B | Deceptive | Holistic | AI-generated videos manipulated to look real and used for malicious purposes. | 24 |
| 93 | 2021 | Al-Dhabi and Zhang | A→B | Deceptive | Holistic | Artificially generated videos that use deep learning to mimic real content for deceptive purposes. | 39 |

| 94 | 2021 | Alheeti et al. | A→B | Non-deceptive | Holistic | Deepfakes are artificially generated images that use techniques like GANs to swap or alter faces in a video. | 8 |
|---|---|---|---|---|---|---|---|
| 95 | 2021 | Ali et al. | A→B | Deceptive | Targeted | Deepfakes are artificially generated images or videos that manipulate identities through techniques like face swapping. | 146 |
| 96 | 2021 | Biswas et al. | A→B | Deceptive | Holistic | Artificial video manipulations used for spreading false information, detected through advanced techniques like 3D-Xception Net with DFT. | 18 |
| 97 | 2021 | Cao and Gong | A→B | Deceptive | Holistic | Deepfakes are artificial face images that mimic real identities through techniques like face swapping and identity impersonation. | 28 |
| 98 | 2021 | Caporusso | A→B | Deceptive | Holistic | Artificially generated fake content that mimics real data accurately, such as images, videos, audio, and text. | 45 |
| 99 | 2021 | Chaudhary et al. | A→B | Deceptive | Targeted and Holistic | Techniques that alter video frames to mimic real-world appearances through methods like face swapping and attribute manipulation. | 9 |
| 100 | 2021 | Chen and Tan | A→B | Deceptive | Holistic | Artificially generated videos with manipulated identities to deceive systems. | 20 |
| 101 | 2021 | Choi | Mixed Types | Deceptive | Targeted | Deepfakes discussed involve image manipulation leading to AI failures, such as face swapping and identity replacement. | 62 |
| 102 | 2021 | Cochran and Napshin | A→B | Deceptive | Targeted | Artificially generated images or videos that realistically swap faces between individuals, often used for deception or identity replacement. | 48 |
| 103 | 2021 | Collins and Ebrahimi | A→B | Deceptive | Holistic | Artificially generated digital content created using machine learning technology that can manipulate faces or other visual elements to appear as if they are real. | 6 |
| 104 | 2021 | Das et al. | A→B | Deceptive | Holistic | Artificially altered videos where faces are swapped to create deceptive identities or expressions. | 22 |
| 105 | 2021 | Perez Dasilva et al. | A→B | Deceptive | Holistic | Artificially generated videos that involve face swapping or revenge porn to impersonate individuals. | 41 |
| 106 | 2021 | Dheeraj et al. | A→B | Deceptive | Holistic | Deepfakes are AI-generated images that can be manipulated to resemble another person's face or expression, often used for deceptive purposes. | 12 |
| 107 | 2021 | Fallis | A→B | Deceptive | Targeted | Artificially generated videos depicting people's actions or statements that are not their own, used to mislead or spread false information. | 249 |
| 108 | 2021 | Floridi | A→B | Deceptive | Targeted | Deepfakes are used to create convincing fake works, particularly in art authentication where they help detect forgeries. | 218 |
| 109 | 2021 | Gülaçtı and Kahraman | Mixed Types | Deceptive and Non-deceptive | Targeted | AI-generated alterations to existing artworks, challenging traditional notions of authorship and creativity. | 20 |
| 110 | 2021 | Godulla et al. | A→B | Deceptive | Holistic | Deepfakes are AI-generated videos that realistically swap or replace faces between individuals, often used for deception or harmful purposes. | 81 |
| 111 | 2021 | Guhagarkar et al. | A→B | Non-deceptive | Holistic | Deepfakes are AI-generated manipulated videos that appear original but are fake. | 7 |
| 112 | 2021 | Hancock and Bailenson | A→B | Deceptive | Holistic | Deepfakes are artificial videos created using machine learning that mimic real identities through face swapping or attribute manipulation. | 185 |
| 113 | 2021 | Hu et al. | A→B | Deceptive | Holistic | Artificially generated videos that use compression techniques to avoid obvious manipulation while appearing realistic. | 182 |
| 114 | 2021 | Iacobucci et al. | A→B | Deceptive | Holistic | Deepfakes are highly realistic AI-generated videos that mimic real content but are designed to be deceptive or harmful. | 57 |
| 115 | 2021a | Ismail et al. | A→B | Deceptive | Holistic | Artificially generated videos that realistically swap faces between people for deception. | 37 |
| 116 | 2021b | Ismail et al. | A→B | Deceptive | Holistic | Artificially generated videos that involve face swapping to deceive or impersonate individuals. | 126 |

| 117 | 2021 | Jha and Jain | A→B | Deceptive | Holistic | Deepfakes are artificial videos created using advanced machine learning algorithms, specifically Generative Adversarial Networks (GANs), that mimic real-life scenarios by altering or replacing specific features of individuals' identities. | 2 |
|---|---|---|---|---|---|---|---|
| 118 | 2021 | Johnson | A→B | Deceptive | Holistic | Deepfakes used to simulate complex geopolitical scenarios involving identity manipulation and counterretaliation dynamics, particularly in the context of potential nuclear escalation. | 30 |
| 119 | 2021 | Kaliyar et al. | A→B | Deceptive | Holistic | Deepfakes used in this paper are tools or methods employed to detect fake news on social platforms. | 189 |
| 120 | 2021 | Karasavva and Noorbhai | A→B | Non-deceptive | Holistic | Algorithmically synthesized material where a face is superimposed onto another body, often used for non-consensual pornography. | 68 |
| 121 | 2021 | Kerner and Risse | A→B | Deceptive | Holistic | Synthetic video manipulation where identities are replaced or altered. | 63 |
| 122 | 2021 | Khalid et al. | A→B | Deceptive | Holistic | Deepfakes are AI-generated audio and video content designed to impersonate real identities. | 267 |
| 123 | 2021 | Khalil and Maged | A→B | Deceptive | Holistic | Deepfakes are artificial images created using deep learning algorithms that mimic real images and can be used for deceptive purposes. | 52 |
| 124 | 2021 | Khanjani et al. | A→B | Deceptive | Holistic | Artificially generated audio content created using AI methods to mimic real sounds or deceive. | 34 |
| 125 | 2021 | Khichi and Yadav | A→B | Deceptive | Holistic | AI-generated forgeries used to deceive by mimicking real identities or voices. | 17 |
| 126 | 2021 | Khochare et al. | Mixed Types | Deceptive and Non-deceptive | Targeted | Manipulated audio data used to deceive in speaker verification systems. | 95 |
| 127 | 2021 | Khormali and Yuan | A→B | Deceptive | Holistic | Deepfakes are synthetic videos created using GANs that appear realistic, often used for deceptive purposes. | 44 |
| 128 | 2021 | Kietzmann et al. | A→A' | Deceptive | Targeted | Artificially generated videos that use AI to superimpose voices and likenesses, effectively altering identities or messages. | 105 |
| 129 | 2021 | Kiselev | A→B | Deceptive | Targeted | Deepfakes are technology-based tools that use generative adversarial networks to alter or swap identities, often for deceptive purposes. | 20 |
| 130 | 2021 | Kocsis | A→B | Deceptive | Targeted and Holistic | Deepfakes are described as technology used to create artificial images or videos that mimic real data sets, often with altered identities. | 26 |
| 131 | 2021 | Kohli and Gupta | A→B | Deceptive and Non-deceptive | Holistic | Artificially generated video contents that manipulate or replace facial identities for deceptive purposes. | 77 |
| 132 | 2021 | Kwok and Koh | Mixed Types | Deceptive | Holistic | AI-generated videos that manipulate or impersonate identities through face swapping or attribute modification. | 171 |
| 133 | 2021 | Langguth et al. | A→B | Deceptive | Targeted | Deepfakes involve image manipulation using AI to alter or swap identities between people, creating realistic but deceptive content. | 61 |
| 134 | 2021 | Lee and Kim | A→B | Deceptive and Non-deceptive | Targeted | Artificially generated videos that use AI technology to create realistic fake content, often involving identity manipulation. | 37 |
| 135 | 2021 | Leibowicz et al. | A→B | Deceptive | Holistic | Deepfakes are artificial synthetic media designed to deceive or mislead about their origin or authenticity. | 30 |
| 136 | 2021 | Liu et al. | A→B | Deceptive | Targeted | Artificially generated videos that involve realistic face swaps or identity modifications to deceive viewers. | 58 |
| 137 | 2021 | Liu et al. | A→B | Deceptive | Holistic | Deepfakes are manipulated images or videos used to deceive individuals or organizations into believing false information. | 6 |
| 138 | 2021 | Mahmud and Sharmin | A→B | Deceptive | Holistic | Artificially generated content that mimics real images or videos using deep learning techniques to manipulate attributes like faces or voices. | 78 |

| 139 | 2021 | Meckel and Steinacker | A→B | Deceptive | Holistic | Deepfakes are synthetic media generated using AI techniques like deep learning to manipulate images or voices. | 9 |
|---|---|---|---|---|---|---|---|
| 140 | 2021 | Meneses | A→B | Deceptive | Targeted and Holistic | Artificially manipulated media used for deception, particularly in political contexts to spread misinformation. | 7 |
| 141 | 2021 | Mihailova | A→B | Deceptive | Holistic | Deepfakes are technologies that create realistic face swaps or identity modifications for artistic and educational purposes. | 50 |
| 142 | 2021 | Mitra et al. | A→B | Deceptive | Holistic | Deepfakes are AI-generated images manipulated to look real through GAN technology. | 15 |
| 143 | 2021 | Mitra et al. | A→B | Deceptive | Holistic | Deepfakes are videos where a person's face, emotion, or speech is replaced with another's using deep learning. | 92 |
| 144 | 2021 | Neethirajan | A→B | Deceptive | Holistic | Deepfakes used here involve generating synthetic data based on real distributions to train models, particularly focusing on animal identities and emotions. | 31 |
| 145 | 2021 | Negi et al. | A→B | Deceptive | Holistic | Artificially generated fake images, audio, and videos that appear realistic but are not genuine. | 21 |
| 146 | 2021 | Nema | A→B | Deceptive | Holistic | AI-generated alterations of media to deceive or replace original content. | 10 |
| 147 | 2021 | Pandey et al. | A→B | Deceptive | Holistic | Deepfakes involve realistic face swapping or identity replacement between individuals. | 12 |
| 148 | 2021 | Pashine et al. | A→B | Deceptive | Holistic | Deepfakes are artificial images or videos with manipulated faces created using neural networks to deceive or impersonate individuals. | 37 |
| 149 | 2021 | Patil and Chouragade | A→B | Deceptive | Holistic | Deepfakes are used to create fake videos that mimic real ones for deceptive purposes. | 13 |
| 150 | 2021 | Patil and Chouragade | A→B | Deceptive | Holistic | Artificially generated videos that manipulate identities through face swapping or impersonation to spread false information. | 7 |
| 151 | 2021 | Patil et al. | A→B | Deceptive | Holistic | Artificially generated videos that impersonate real individuals or events using deep learning techniques. | 9 |
| 152 | 2021 | Pavis | A→B | Deceptive | Holistic | Synthetic performances created using AI systems, often involving face swapping or identity replacement. | 38 |
| 153 | 2021 | Qureshi et al. | A→B | Deceptive | Holistic | Artificially generated videos or audio created using AI algorithms to deceive or impersonate individuals. | 35 |
| 154 | 2021 | Rafique et al. | A→B | Deceptive | Holistic | Deepfakes are realistic face swaps used for deception. | 48 |
| 155 | 2021 | Rana et al. | A→B | Deceptive | Targeted | Artificially generated videos that manipulate identities through face swapping or similar techniques to deceive. | 25 |
| 156 | 2021 | Rao et al. | A→B | Deceptive | Targeted | Deepfakes used in this context are artificial images or videos that impersonate real people to spread false information. | 202 |
| 157 | 2021 | Rashid et al. | A→B | Deceptive | Holistic | Deepfakes are artificial manipulations of digital content used to deceive or misrepresent facts. | 36 |
| 158 | 2021 | Ratner | A→B | Deceptive | Holistic | AI-generated images or videos used to create realistic child abuse scenarios without actual abuse occurring. | 28 |
| 159 | 2021 | Reid | A→B | Deceptive | Targeted and Holistic | Artificially generated videos that realistically swap faces between individuals. | 36 |
| 160 | 2021 | Revi et al. | A→B | Deceptive and Non-deceptive | Targeted | Deepfakes are AI-generated images or videos used to deceive by impersonating individuals or entities. | 34 |
| 161 | 2021 | Ruiter | A→B | Deceptive | Holistic | Deepfakes involve the creation of realistic video or audio representations that misrepresent individuals' identities or expressions. | 172 |
| 162 | 2021 | Rupapara et al. | A→B | Deceptive | Holistic | Deepfakes in this context refer to fake content generated using deep learning models like stacked Bi-LSTM networks for tasks such as tweet classification. | 41 |
| 163 | 2021 | Saeedi et al. | A→B | Deceptive | Targeted | AI-generated misleading content used for deception or harm in consumer applications. | 19 |

| 164 | 2021 | Saravani et al. | A→B | Non-deceptive | Holistic | Deepfakes used for detecting bots by analyzing their text content to identify patterns or mimic human-like language. | 20 |
|---|---|---|---|---|---|---|---|
| 165 | 2021 | Seta | A→A' | Deceptive | Targeted | Deepfakes are synthetic media created by altering or swapping faces in images or videos to create deceptive content. | 30 |
| 166 | 2021 | Shende et al. | A→B | Deceptive | Holistic | Deepfakes are AI-generated videos that use realistic face swapping to create appearances of real identities, often for deceptive purposes. | 12 |
| 167 | 2021 | Su et al. | A→B | Deceptive | Targeted and Holistic | Artificially generated videos that involve tampering or alteration of facial features using AI techniques. | 22 |
| 168 | 2021 | Tahir et al. | A→B | Deceptive | Holistic | AI-generated videos that manipulate real-world identities to appear authentic for deceptive purposes. | 72 |
| 169 | 2021 | Taylor | A→B | Deceptive | Holistic | Deepfakes are artificial face swapping technologies used to replace identities in social media and political contexts. | 32 |
| 170 | 2021 | Thambawita et al. | Mixed Types | Deceptive | Holistic | Synthetic data generated to represent real medical data such as ECGs for privacy reasons. | 4 |
| 171 | 2021 | Thambawita et al. | A→A' | Deceptive | Holistic | Synthetic ECGs generated using GANs to mimic real data for privacy purposes. | 91 |
| 172 | 2021 | Truby and Brown | A→B | Non-deceptive | Holistic | Deepfakes in this context are synthetic representations or clones that mimic human digital behavior to enhance AI's ability to predict and target consumers accurately. | 43 |
| 173 | 2021 | Tu et al. | A→B | Deceptive | Holistic | Inferred from context as technologies used for deceptive manipulations like face swapping and identity replacement leading to harm. | 5 |
| 174 | 2021 | Tuomi | A→B | Deceptive | Holistic | Synthetic reviews designed to deceive consumers or businesses, often used for legitimate business advantages without malicious intent. | 21 |
| 175 | 2021 | Ullrich | A→B | Deceptive | Targeted | Artificially generated videos that substitute one person's likeness with another to deceive viewers. | 26 |
| 176 | 2021 | Verma et al. | A→B | Deceptive | Holistic | Deepfakes are artificial face manipulations detected using Inception-ResnetV2 for security purposes. | 4 |
| 177 | 2021 | Vizoso et al. | A→B | Deceptive | Targeted | AI-enhanced manipulations such as face swapping, sound alterations, or photo doctoring used to spread misinformation. | 86 |
| 178 | 2021 | Volodin et al. | A→B | Deceptive and Non-deceptive | Holistic | Artificially generated face swaps for deceptive purposes in entertainment and social media contexts. | 3 |
| 179 | 2021 | Volodin et al. | A→B | Deceptive | Holistic | Artificially generated content that manipulates input data to hide or alter information, specifically through image compression to remove extraneous noise. | 3 |
| 180 | 2021 | Wahl-Jorgensen and Carlson | A→B | Deceptive | Targeted | Deepfakes are artificial videos that manipulate faces or identities to create deceptive content. | 66 |
| 181 | 2021 | Whittaker et al. | A→B | Deceptive | Holistic | Deepfakes are AI-generated digital content that involves realistic alterations to images or videos, often used to replace or modify identities. | 84 |
| 182 | 2021 | Wiederhold | A→A' | Deceptive | Targeted | Synthetic media created using AI and ML to create convincing images, videos, or audio that mimic real-world environments or people. | 13 |
| 183 | 2021 | Wilkerson | A→B | Deceptive | Holistic | Artificially altered videos used to damage reputations or identities, often for deceptive purposes. | 39 |
| 184 | 2021 | Xia and Hua | A→B | Deceptive | Holistic | Deepfakes are realistic video manipulations that involve swapping or altering faces to deceive viewers. | 2 |
| 185 | 2021 | Yadav et al. | A→B | Deceptive | Holistic | AI-generated videos where target images replace source ones for deceptive purposes. | 13 |
| 186 | 2021 | Yadlin-Segal and Oppenheim | A→B | Deceptive | Holistic | AI-based manipulations used for spreading disinformation or misleading purposes. | 116 |
| 187 | 2021 | Zhang et al. | A→B | Deceptive | Holistic | Deepfakes are photorealistic synthetic videos used to deceive or mislead by altering identities or expressions. | 85 |

| 188 | 2021 | Zhao et al. | A→B | Deceptive | Targeted and Holistic | Artificially generated visual representations that manipulate or alter geospatial data to create false landscapes or features. | 146 |
|---|---|---|---|---|---|---|---|
| 189 | 2021 | Zhao et al. | Mixed Types | Deceptive | Holistic | AI-generated videos with manipulated identities used for deceptive purposes. | 28 |
| 190 | 2021 | Zhao et al. | A→B | Deceptive | Holistic | Synthetic videos with altered identities created using AI tools like FaceSwap and StyleGAN. | 9 |
| 191 | 2021 | Zobaed et al. | A→B | Deceptive | Targeted and Holistic | Synthetic images or videos with altered identities created using deep learning techniques to deceive or manipulate. | 40 |
| 192 | 2022 | Abdulreda and Obaid | A→B | Deceptive and Non-deceptive | Holistic | Techniques involving deep learning to detect or create manipulated images, often focusing on face swapping for identity transfer. | 3 |
| 193 | 2022 | Abu-Ein et al. | A→B | Deceptive | Holistic | Artificially generated digital content that realistically alters or replaces parts of an image or video to deceive viewers. | 16 |
| 194 | 2022 | Adnan and Abdulbaqi | A→B | Deceptive | Holistic | Artificially generated videos with manipulated faces used for detection purposes. | 7 |
| 195 | 2022 | Ahmed et al. | A→B | Deceptive | Holistic | Deepfakes are artificial images or videos created using deep learning techniques that mimic real content for deceptive purposes. | 175 |
| 196 | 2022 | Alanazi | Mixed Types | Deceptive | Holistic | Deepfakes are artificial videos that use deep learning to create realistic fake images or modify existing ones, often for deceptive purposes. | 6 |
| 197 | 2022a | Ali et al. | A→B | Deceptive | Targeted and Holistic | Deepfakes are AI-generated deep learning-based technologies that allow realistic identity swapping or impersonation. | 15 |
| 198 | 2022b | Ali et al. | A→B | Deceptive and Non-deceptive | Holistic | Deepfakes involve altering digital media such as faces, images, or movies to deceive individuals or organizations for malicious purposes like fraud or identity replacement. | 25 |
| 199 | 2022 | Almutairi and Elgibreen | A→B | Deceptive | Holistic | Artificially generated audio manipulations used for deception. | 143 |
| 200 | 2022 | Altaei | Mixed Types | Deceptive | Targeted and Holistic | Artificially generated face images that mimic real ones for deceptive purposes. | 25 |
| 201 | 2022 | Amezaga and Hajek | A→B | Deceptive | Targeted | Artificially generated voices that resemble human speech to impersonate others. | 18 |
| 202 | 2022 | Appel and Prietzel | A→B | Deceptive | Targeted | AI-generated manipulations used for political purposes, often involving face swapping or attribute changes. | 106 |
| 203 | 2022 | Ataş et al. | A→B | Deceptive | Holistic | Artificially generated videos that involve alterations in identity or attributes to deceive users. | 12 |
| 204 | 2022 | İlhan et al. | A→B | Deceptive and Non-deceptive | Holistic | Deepfakes are AI-generated images that mimic real ones, often used for fraudulent purposes. | 9 |
| 205 | 2022 | Awotunde et al. | A→B | Deceptive | Holistic | Artificially generated videos that manipulate faces or identities to appear genuine. | 31 |
| 206 | 2022 | Balasubramanian et al. | A→B | Deceptive | Holistic | AI-generated fake images or videos used to deceive. | 41 |
| 207 | 2022 | Barabanshchikov and Marinova | A→B | Deceptive | Holistic | A technology enabling the creation of video clips with manipulated or swapped faces to create 'impossible' images without obvious alterations. | 4 |
| 208 | 2022 | Bateman | A→B | Deceptive | Holistic | Artificially generated content used for malicious purposes such as identity impersonation, financial fraud, and market manipulation. | 98 |
| 209 | 2022 | Campbell et al. | A→B | Deceptive and Non-deceptive | Holistic | Deepfakes in this context refer to AI-generated manipulations of advertising content, such as synthetic ads with altered or swapped faces. | 275 |
| 210 | 2022 | Campbell et al. | A→B | Deceptive | Holistic | AI-driven tools used to create convincing fake identities or images, particularly through face swapping. | 91 |

| 211 | 2022 | Concas et al. | Mixed Types | Deceptive | Targeted and Holistic | Deepfakes are artificially generated images that manipulate faces using GAN-based algorithms to create realistic fake identities. | 20 |
|---|---|---|---|---|---|---|---|
| 212 | 2022 | Concas et al. | A→B | Deceptive and Non-deceptive | Holistic | Deepfakes are artificial alterations of visual identities used for deceptive purposes in detection systems. | 15 |
| 213 | 2022 | Cross | A→B | Non-deceptive | Holistic | Deepfakes are used to create convincing fake identities through AI, often involving face swapping or identity replacement. | 45 |
| 214 | 2022 | Custers | A→B | Deceptive | Targeted | Deepfakes used for evaluating the effectiveness of sanctions and law enforcement tools, such as A/B optimization and algorithmic profiling in criminal law. | 37 |
| 215 | 2022 | Das et al. | A→B | Deceptive | Holistic | Deepfakes are artificially generated videos that use techniques like face swapping, altered facial expressions, gender change, fake content creation, and altered facial features to create realistic but manipulated images or sequences. | 13 |
| 216 | 2022 | Delfino | A→B | Deceptive | Targeted | Artificially generated audiovisual recordings that map one person's movements and words onto another to create realistic identities. | 62 |
| 217 | 2022 | Deng et al. | A→B | Deceptive | Holistic | Tampered images or videos where faces are altered or replaced, leaving synthetic traces at edges. | 5 |
| 218 | 2022 | Eberl et al. | A→B | Deceptive | Holistic | Deepfakes are artificial data generated using deep learning to simulate or alter human identities in images or videos. | 31 |
| 219 | 2022 | Galyashina and Nikishin | A→A' | Deceptive | Holistic | Deepfakes are artificial voice clones used for intellectual or material forgery. | 9 |
| 220 | 2022 | Ganguly et al. | A→B | Deceptive | Holistic | Deepfakes are synthetic videos where faces or identities are manipulated to replace real ones using computer graphics and vision techniques. The paper focuses on detecting such manipulations. | 35 |
| 221 | 2022 | Ge et al. | A→B | Deceptive | Holistic | Deepfakes are technologies used to create deceptive representations of individuals for authentication or impersonation purposes. | 45 |
| 222 | 2022 | Gowda and Thillaiarasu | A→B | Deceptive | Holistic | Deepfakes are AI-generated videos that use techniques like GANs and CNNs to swap or modify faces, making them appear real. | 12 |
| 223 | 2022 | Gu et al. | A→B | Deceptive | Targeted | Artificially generated videos that involve realistic face manipulation to deceive or impersonate individuals. | 104 |
| 224 | 2022 | Gu et al. | A→B | Deceptive | Holistic | Video-based deepfakes that involve face forgery and manipulation to deceive or impersonate individuals. | 25 |
| 225 | 2022 | Guarnera et al. | A→B | Deceptive | Holistic | Artificially generated images or videos that manipulate faces to deceive. | 58 |
| 226 | 2022 | Guzman | A→B | Deceptive | Holistic | Artificially generated videos depicting realistic depictions of intimate scenes without consent, often used for harmful purposes. | 9 |
| 227 | 2022 | Hadi et al. | Mixed Types | Deceptive | Targeted | Deepfakes involve the creation of artificial video or image content where parts of a person's identity are replaced or modified using AI techniques like GANs. | 3 |
| 228 | 2022 | Hamza et al. | A→B | Deceptive | Targeted and Holistic | Artificial audio manipulation using MFCC features to mimic real sounds for detection purposes. | 143 |
| 229 | 2022 | Hanif and Dave | A→B | Deceptive | Targeted | Deepfakes are computerized methods to create counterfeit content that is difficult for humans to distinguish from real media. | 8 |
| 230 | 2022 | Hao et al. | A→B | Deceptive | Holistic | Artificially generated content that involves face manipulation for deceptive purposes. | 21 |
| 231 | 2022 | Huang et al. | A→B | Deceptive | Targeted | Artificially generated images that manipulate facial attributes or entire faces to deceive individuals. | 112 |
| 232 | 2022 | Humidan et al. | A→B | Deceptive | Holistic | Synthetic video or images used for deception and spreading harmful content. | 5 |

| 233 | 2022 | Ismail et al. | A→B | Deceptive | Holistic | Deepfakes are videos where faces are swapped between individuals to create realistic-looking fake content, often used for malicious purposes like identity theft or blackmail. | 23 |
|---|---|---|---|---|---|---|---|
| 234 | 2022 | Jaleel and Ali | A→B | Deceptive | Holistic | Artificially generated videos that mimic real identities through detailed facial analysis. | 10 |
| 235 | 2022 | Jameel et al. | A→B | Deceptive | Holistic | Artificially generated images that manipulate identities through face swapping or attribute replacement. | 9 |
| 236 | 2022 | Jayakumar and Skandhakumar | Mixed Types | Deceptive | Holistic | Deepfakes are videos manipulated using AI techniques to alter identities, often through face swapping or attribute modification. | 13 |
| 237 | 2022 | Jeong et al. | A→B | Deceptive | Targeted | A method to detect and identify deepfakes by analyzing frequency-level perturbations in generated images. | 98 |
| 238 | 2022 | John and Sherif | A→B | Deceptive | Targeted and Holistic | Deepfakes involve tampered images or videos where a face is superimposed to deceive. | 16 |
| 239 | 2022 | Joost | A→B | Deceptive | Holistic | Deepfakes are artificial images or videos that impersonate real people, often used for deceptive purposes like identity theft or fraud. | 14 |
| 240 | 2022 | Kandasamy et al. | A→B | Deceptive | Holistic | Artificially generated images or videos used for spreading misinformation, often computationally created using deep learning models like LSTM and CNNs. | 15 |
| 241 | 2022 | Kasita | A→B | Deceptive and Non-deceptive | Holistic | Artificially generated content, such as images or videos, that manipulate identities for harmful purposes. | 26 |
| 242 | 2022 | Kaushal et al. | A→A' | Deceptive | Targeted | Deepfakes are artificial images or videos created using deep learning to appear realistic. | 5 |
| 243 | 2022 | Kawa et al. | A→A' | Deceptive | Targeted | Deepfakes used in this context are synthetic audio recordings designed to deceive listeners by mimicking genuine speech or sounds. | 12 |
| 244 | 2022 | Khan et al. | A→B | Deceptive and Non-deceptive | Holistic | Artificially generated audio that mimics real speech using ML algorithms to detect and distinguish it from genuine audio. | 5 |
| 245 | 2022 | Khoo et al. | A→B | Deceptive and Non-deceptive | Targeted | Artificially generated images that mimic real-world content with high fidelity. | 41 |
| 246 | 2022 | Kim et al. | A→B | Deceptive | Targeted | Deepfakes created using advanced algorithms to manipulate medical images for fraudulent purposes. | 7 |
| 247 | 2022 | Kirn et al. | A→B | Deceptive | Holistic | Artificially generated fake tweets that involve altering user identities or content to deceive. | 10 |
| 248 | 2022 | KoÇak and Alkan | A→B | Deceptive | Holistic | Artificially generated images or videos created using deep learning to mimic real content, often used for deception. | 8 |
| 249 | 2022 | Kraetzer et al. | A→B | Deceptive | Holistic | Artificially generated media with altered identities or attributes used for detection purposes. | 12 |
| 250 | 2022 | Kshirsagar et al. | Mixed Types | Deceptive | Targeted | Artificially generated content aiming for deception or authenticity manipulation. | 9 |
| 251 | 2022 | Kumar and Alraisi | A→B | Deceptive | Holistic | Not explicitly defined but used for malicious purposes like identity theft. | 12 |
| 252 | 2022 | López-Gil | A→B | Deceptive | Holistic | Deepfakes are videos that use artificial intelligence and machine learning to create fake images or identities which can be used to spread false information, particularly for political purposes. | 11 |
| 253 | 2022 | Lübbeling | A→B | Deceptive | Holistic | Artificially generated videos that realistically alter or replace individuals' identities through face swapping or voice manipulation. | 2 |
| 254 | 2022 | Lacerda and Vasconcelos | A→B | Non-deceptive | Holistic | Deepfakes are images or videos where faces are manipulated to appear authentic but aren't real, often created using AI techniques like convolutional neural networks. | 4 |

| 255 | 2022 | Lalitha and Sooda | Mixed Types | Deceptive | Targeted | Artificially manipulated videos where faces, emotions, or speech are substituted to deceive. | 19 |
|---|---|---|---|---|---|---|---|
| 256 | 2022 | Lee et al. | A→B | Deceptive | Holistic | Synthetic images designed to deceive detection systems by altering or replacing identities, particularly using facemasks. | 9 |
| 257 | 2022 | Lewis | A→A' | Deceptive | Targeted | Artificially generated content used for deception or harmful purposes. | 7 |
| 258 | 2022 | Lewis et al. | A→B | Deceptive | Holistic | AI-generated videos manipulated to look real, often involving face swapping or realistic alterations. | 13 |
| 259 | 2022 | Li et al. | A→B | Deceptive | Holistic | Artificial audio signals manipulated to appear as real recordings, often using deep learning techniques. | 30 |
| 260 | 2022 | Lim et al. | A→B | Deceptive | Holistic | Deepfakes used in adversarial attacks to deceive detection models, often involving face swapping or similar manipulations to mimic another person's identity. | 12 |
| 261 | 2022 | Lim et al. | A→B | Deceptive | Holistic | Artificially generated audio data that manipulates voice characteristics to impersonate individuals. | 69 |
| 262 | 2022 | Liu and Zhang | A→B | Deceptive and Non-deceptive | Holistic | Deepfakes are AI-generated fake content created by splicing individual sounds, facial expressions, and body movements into false content using neural network technology. | 11 |
| 263 | 2022 | Liu et al. | A→B | Deceptive and Non-deceptive | Holistic | Artificially generated images that involve swapping or altering faces to create fake identities. | 3 |
| 264 | 2022 | Lorch et al. | A→B | Deceptive | Holistic | Artificially generated images used to replace or deceive identities in law enforcement contexts such as fraud detection and license plate recognition. | 16 |
| 265 | 2022 | Lucas | A→B | Deceptive | Targeted | Artificially generated video content that manipulates faces and situations to misrepresent identities or events. | 45 |
| 266 | 2022 | Müller et al. | A→B | Deceptive | Holistic | Implicitly, deepfakes involve audio manipulation to imitate someone's voice. | 83 |
| 267 | 2022 | Malik et al. | A→B | Deceptive | Holistic | DeepFakes are synthetic face images or videos created using deep learning algorithms to replace or alter facial features in a realistic manner. | 192 |
| 268 | 2022 | Mallet et al. | A→B | Deceptive | Holistic | Artificially generated videos that replace one person's face with a more recognizable individual to spread misinformation or achieve malicious goals. | 22 |
| 269 | 2022 | Matthews | A→B | Deceptive | Holistic | Artificially generated videos depicting people doing and saying things they never did. | 13 |
| 270 | 2022 | Mishra and Samanta | A→B | Deceptive | Holistic | Deepfakes are artificially generated images that manipulate faces to deceive or impersonate individuals. | 8 |
| 271 | 2022 | Mullen | A→B | Deceptive | Holistic | Deepfakes are synthetic images or videos that mimic real people's identities through realistic features like facial expressions. | 20 |
| 272 | 2022 | Myvizhi and Pamila | Mixed Types | Deceptive | Holistic | Deepfakes are manipulated videos where identities or features are altered using deep learning techniques to create realistic fake content. | 10 |
| 273 | 2022 | Nagothu et al. | A→B | Deceptive | Holistic | Artificially generated audio or video content that mimics real content to deceive. | 16 |
| 274 | 2022 | Nawaz et al. | A→B | Deceptive | Holistic | Deepfakes are artificially generated videos that use techniques like FaceSwap to swap faces between individuals, often for deceptive purposes. | 8 |
| 275 | 2022 | Nelson and Lewis | A→B | Deceptive and Non-deceptive | Holistic | Artificially generated videos that realistically swap faces between people. | 7 |
| 276 | 2022 | Nguyen et al. | A→B | Deceptive | Holistic | Deepfakes are AI-generated fake images and videos that mimic real content to deceive viewers. | 625 |
| 277 | 2022 | Nour and Gelfand | A→B | Deceptive | Holistic | Artificially generated audiovisual content that combines deep learning algorithms with enhancements to create fake or misleading information, often used for misinformation. | 7 |

| 278 | 2022 | Park et al. | A→B | Deceptive | Holistic | Artificially generated videos that manipulate movements of participants in real-time. | 7 |
|---|---|---|---|---|---|---|---|
| 279 | 2022 | Patel et al. | A→B | Deceptive | Targeted | Artificially generated fake videos that use AI to create realistic content such as face swaps or manipulated identities. | 7 |
| 280 | 2022 | Pipin et al. | A→B | Deceptive | Holistic | Deepfakes are videos that use techniques like face swapping and photo response non-uniformity analysis to create or detect fake content aimed at misleading consumers. | 13 |
| 281 | 2022 | Pooyandeh et al. | A→B | Deceptive | Holistic | Not explicitly defined but inferred as part of cybersecurity discussions in an AI-based Metaverse context. | 134 |
| 282 | 2022 | Preu et al. | A→B | Deceptive and Non-deceptive | Targeted | Synthetic media used for deception or misleading purposes. | 12 |
| 283 | 2022 | Prezja et al. | A→B | Deceptive | Holistic | Synthetic images created by generative adversarial networks (GANs) that aim to mimic real medical X-ray images of knees with varying degrees of osteoarthritis. | 54 |
| 284 | 2022 | Pu et al. | A→B | Deceptive | Targeted and Holistic | DeepFakes are artificial face manipulations where identities are swapped or modified to deceive. | 56 |
| 285 | 2022 | Purwadi et al. | A→B | Deceptive | Holistic | Artificial manipulation of digital personal data to create false identities or images for malicious purposes. | 8 |
| 286 | 2022 | Rahman et al. | A→B | Deceptive | Holistic | Artificially generated videos that appear real but are manipulated to deceive or impersonate individuals. | 19 |
| 287 | 2022 | Ramkissoon et al. | A→B | Deceptive | Holistic | DeepFakes are videos manipulated using realistic techniques to alter or impersonate identities. | 3 |
| 288 | 2022 | Rana et al. | A→B | Deceptive | Holistic | AI-generated deepfakes used for deception and manipulation in various applications. | 393 |
| 289 | 2022 | Rao et al. | A→B | Deceptive | Holistic | AI-generated videos that are fabricated to appear real but are actually fake. | 7 |
| 290 | 2022 | Raza et al. | A→B | Deceptive | Holistic | Deepfakes are artificial images or videos that manipulate real identities to deceive users. | 123 |
| 291 | 2022 | Rini and Cohen | A→B | Deceptive | Holistic | Digitally altered audio or video recordings where one person's face and/or voice is replaced with another's. | 71 |
| 292 | 2022 | Saadi and Al-Jawher | A→B | Deceptive | Holistic | Artificially generated videos where faces are manipulated or swapped to create fake content. | 11 |
| 293 | 2022 | Saber et al. | A→B | Deceptive | Holistic | Deepfakes are artificial face-swapping techniques used to create fake videos that appear legitimate. | 3 |
| 294 | 2022 | Saikia et al. | A→B | Deceptive | Holistic | Synthetic video content created using deep learning algorithms to mimic real-life scenarios. | 90 |
| 295 | 2022 | Salini and HariKiran | A→B | Deceptive | Holistic | Synthetic images generated by GANs that appear authentic and can be indistinguishable from real fundus images. | 8 |
| 296 | 2022 | Saxena and KN | A→B | Deceptive | Holistic | Deepfakes are images where the face of one person is replaced with another's to create a realistic or misleading appearance. | 18 |
| 297 | 2022 | Seow et al. | A→B | Deceptive and Non-deceptive | Holistic | Artificially generated images or videos that manipulate faces or identities for various purposes. | 124 |
| 298 | 2022 | Shahzad et al. | A→B | Deceptive | Holistic | Artificially generated images and videos that manipulate identities using AI to deceive. | 80 |
| 299 | 2022 | Shin and Lee | A→B | Deceptive | Holistic | Artificially generated videos that realistically swap faces between people to create false identities or scenarios. | 51 |
| 300 | 2022 | Siegel et al. | A→B | Deceptive and Non-deceptive | Holistic | AI-generated manipulations used for fraudulent purposes, such as face swapping to deceive. | 14 |
| 301 | 2022 | Slater and Rastogi | A→B | Deceptive | Holistic | Deepfakes are artificial images or videos that use compositing to create realistic synthetic media, often used for deceptive purposes in political contexts. | 3 |

| 302 | 2022 | Sonkusare et al. | A→B | Deceptive | Holistic | Techniques that bypass facial features to deceive biometric authentication systems, often by replacing genuine identities with synthetic ones. | 4 |
|---|---|---|---|---|---|---|---|
| 303 | 2022 | Taeb and Chi | A→B | Deceptive | Holistic | Artificially generated deepfakes that swap or modify faces in video and digital content to deceive. | 89 |
| 304 | 2022 | Taha et al. | A→B | Deceptive | Targeted | Deep fakes are synthetic methods using AI to create manipulated content for deception. | 7 |
| 305 | 2022 | Tran et al. | A→B | Deceptive | Holistic | Deepfakes are artificial face manipulations used to deceive or impersonate individuals. | 19 |
| 306 | 2022 | Vamsi et al. | A→B | Deceptive | Holistic | AI-generated videos that manipulate real identities for deception. | 56 |
| 307 | 2022 | Vera | A→B | Deceptive and Non-deceptive | Targeted | Deepfakes are artificially generated videos that manipulate images or video content to appear realistic. | 2 |
| 308 | 2022 | Vinay et al. | A→B | Deceptive | Holistic | Deepfakes are artificial videos or images created using GAN technology that mimic real identities or scenarios. | 6 |
| 309 | 2022 | Vo et al. | Mixed Types | Deceptive | Targeted | Deepfakes are AI-generated images or videos that mimic real content, particularly focusing on face swapping and identity impersonation. | 6 |
| 310 | 2022 | Wang and Kim | A→B | Deceptive | Targeted and Holistic | AI-generated hyper-realistic manipulated media content, particularly involving face swapping or identity replacement. | 27 |
| 311 | 2022 | Wang et al. | A→B | Deceptive | Targeted | Deepfakes are technologies used to fabricate or manipulate images for fraudulent purposes, such as altering research outputs. | 34 |
| 312 | 2022 | Wang et al. | A→B | Deceptive | Holistic | Artificially generated images with synthetic faces that include added noise to mimic real faces. | 25 |
| 313 | 2022 | Waqas et al. | A→B | Deceptive | Holistic | Deepfakes are synthetic images created by transferring important features from a source image to a target image in a way that the target image appears realistic and resembles the source almost in reality. | 32 |
| 314 | 2022 | Wolter et al. | A→B | Deceptive | Targeted and Holistic | Synthetic images created to mimic real ones, often used for deception or misrepresentation. | 49 |
| 315 | 2022 | Woo | A→B | Deceptive | Holistic | Artificially generated images that manipulate or alter facial identities. | 91 |
| 316 | 2022 | Yan et al. | A→B | Deceptive | Targeted | Deepfakes in this context refer to audio generated by specific tools or models that can be attributed to identify their source. | 3 |
| 317 | 2022 | Yang et al. | A→B | Deceptive | Holistic | Deepfakes are artificial images generated using GANs that deceive by imitating real identities or scenes. | 69 |
| 318 | 2022 | Yu et al. | A→B | Deceptive | Holistic | Deepfakes involve the use of deep learning to replace or alter faces in images or videos, often for deceptive purposes. | 21 |
| 319 | 2022 | Zhang | A→B | Deceptive | Holistic | Artificially generated realistic images, sounds, or videos created using AI methods to deceive. | 158 |
| 320 | 2023 | Łabuz | A→B | Deceptive | Targeted and Holistic | Deepfakes are artificially generated videos that realistically swap faces between people, often used to deceive or mislead. | 13 |
| 321 | 2023 | Abir et al. | A→B | Deceptive | Holistic | Deepfakes are images or videos where one person's identity is replaced with another using advanced AI techniques. | 32 |
| 322 | 2023 | Aggarwal et al. | A→B | Deceptive | Holistic | Deepfakes are images where identities or parts of individuals have been altered or replaced, often for deceptive purposes. | 31 |
| 323 | 2023 | Ahmed and Shaun | A→B | Non-deceptive | Holistic | AI-generated fake videos altering faces with expressions and lip movements. | 3 |
| 324 | 2023 | Aissani et al. | A→B | Deceptive and Non-deceptive | Holistic | In this context, deepfakes refer to false or misleading content created using AI tools, particularly in media and journalism for purposes like generating fake news. | 32 |
| 325 | 2023 | Akhtar | A→B | Deceptive | Targeted and Holistic | Artificially generated digital face content that can be manipulated or synthesized to deceive or alter identities. | 101 |

| 326 | 2023 | Alanazi et al. | A→B | Deceptive | Holistic | Artificially generated images that manipulate facial features to create realistic identities or expressions. | 6 |
|---|---|---|---|---|---|---|---|
| 327 | 2023 | Alkishri and Al-Bahri | Mixed Types | Deceptive | Targeted | Deepfakes are images created using deep learning models that can be mistaken for real content, often used for fraudulent purposes. | 10 |
| 328 | 2023 | Altaei | A→A' | Deceptive | Targeted | Deepfakes are face images where identities have been swapped or replaced to deceive, often used for fraud and identity theft. | 14 |
| 329 | 2023 | Altalahin et al. | A→B | Deceptive | Holistic | Artificially generated audio that appears real using AI techniques to manipulate voice features. | 26 |
| 330 | 2023 | Anagha et al. | A→B | Deceptive and Non-deceptive | Holistic | Synthetic audio recordings designed to mimic real voices or sounds but are not authentic. | 13 |
| 331 | 2023 | Arshed et al. | A→B | Deceptive | Holistic | Artificially generated content that realistically manipulates identities or expressions of individuals. | 13 |
| 332 | 2023 | Arslan | A→B | Deceptive | Holistic | Artificially generated videos that realistically transfer faces or voices between individuals, often used for deception. | 12 |
| 333 | 2023 | Asha et al. | A→B | Non-deceptive | Holistic | Deepfakes are artificial images or videos that manipulate facial traits to create realistic-looking fake identities. | 12 |
| 334 | 2023 | Bansal et al. | A→B | Non-deceptive | Holistic | Artificially altered multimedia content designed to deceive or mislead users. | 8 |
| 335 | 2023a | Beerbaum | A→B | Deceptive | Targeted and Holistic | Deepfakes are discussed as tools used to create realistic fake videos or images that can be manipulated for malicious purposes such as political manipulation or spreading misinformation. | 27 |
| 336 | 2023b | Beerbaum | A→B | Deceptive and Non-deceptive | Holistic | The paper discusses the generation of false or misleading content, such as deepfakes, which are used for harmful purposes like political manipulation and spreading misinformation. | 7 |
| 337 | 2023 | Berrahal et al. | A→B | Deceptive | Holistic | Deepfakes involve the use of artificial intelligence to generate fake images or videos that mimic real ones. | 8 |
| 338 | 2023 | Bilika et al. | A→B | Deceptive | Holistic | Artificial voice deepfakes used to deceive users by impersonating owners of voice assistants. | 14 |
| 339 | 2023 | Bird and Lotfi | A→B | Deceptive | Holistic | AI-generated audio that mimics real human speech to mimic or alter identities. | 32 |
| 340 | 2023 | Boutadjine et al. | A→B | Deceptive | Holistic | Artificially generated media that involves face swapping or identity replacement using deep learning technology to deceive. | 10 |
| 341 | 2023 | BR et al. | A→B | Deceptive and Non-deceptive | Targeted and Holistic | Artificially generated videos that use AI and machine learning to manipulate or superimpose images or audio onto existing footage in a way that appears authentic. | 24 |
| 342 | 2023 | Bray et al. | A→B | Deceptive | Holistic | Deepfakes are uncurated outputs from StyleGAN2 trained on the FFHQ dataset, used to create images of human faces that falsely represent reality. | 80 |
| 343 | 2023 | Bu et al. | A→B | Deceptive | Holistic | Deepfakes are advanced digital manipulations using artificial intelligence to alter or create realistic identities through techniques such as video face changing, audio synthesis, and text forgery. | 5 |
| 344 | 2023 | Busch et al. | A→B | Deceptive | Holistic | Synthetic video content used for malicious purposes such as face swapping to target individuals. | 13 |
| 345 | 2023 | Byman et al. | A→B | Deceptive | Targeted | Artificially generated content used to deceive or manipulate others. | 26 |
| 346 | 2023 | Byman et al. | A→B | Deceptive | Holistic | AI technology used to create realistic images, videos, or text of individuals for deceptive purposes. | 26 |
| 347 | 2023 | Chang et al. | A→B | Deceptive and Non-deceptive | Holistic | AI-generated face manipulations used to deceive or impersonate individuals. | 13 |
| 348 | 2023 | Cheres and Groza | A→B | Deceptive | Holistic | Deepfakes used in this paper are artistic tools that involve face swapping to create alter egos for users. | 5 |

| 349 | 2023 | Chhabra et al. | A→B | Deceptive | Holistic | Deepfakes are manipulated digital media that appear real but involve alterations not easily noticeable by humans. | 10 |
|---|---|---|---|---|---|---|---|
| 350 | 2023 | Choi and Kim | A→B | Non-deceptive | Holistic | Deepfakes are manipulated videos distributed for deceptive purposes such as political attacks or identity theft. | 11 |
| 351 | 2023 | Chong et al. | A→B | Deceptive | Holistic | Deepfakes in this context refer to machine-generated text designed to mimic human-written text for deceptive purposes. | 14 |
| 352 | 2023 | Cocchi et al. | A→B | Deceptive | Holistic | Deepfakes involving image transformations used to bypass detection systems for deceptive purposes. | 17 |
| 353 | 2023 | Coiera et al. | A→B | Deceptive | Holistic | Deepfakes are technologies that use AI models, such as large language models (LLMs), to generate realistic or manipulated visual and textual content, often for creating synthetic identities or images. | 19 |
| 354 | 2023 | Das et al. | A→B | Deceptive | Holistic | Deep fakes are AI-generated fake images or videos created using GANs and autoencoders that replace real identities with synthetic ones. | 7 |
| 355 | 2023 | Dhanaraj and Sridevi | Mixed Types | Deceptive | Targeted | AI-generated videos with face warping used for deception. | 5 |
| 356 | 2023 | Dhesi et al. | A→B | Deceptive | Holistic | Artificially generated images or videos that manipulate or replace facial identities to deceive. | 12 |
| 357 | 2023 | Dixit et al. | A→B | Deceptive | Holistic | Artificially generated audio content that manipulates voice features to impersonate individuals. | 38 |
| 358 | 2023 | Doss et al. | A→B | Deceptive | Holistic | Realistic digital content alterations used to mislead individuals about information sources, particularly in educational contexts. | 47 |
| 359 | 2023 | Dragar et al. | A→B | Deceptive and Non-deceptive | Holistic | AI-generated videos that mimic real identities through realistic face swapping or attribute manipulation to appear authentic. | 3 |
| 360 | 2023 | Durães et al. | A→B | Deceptive and Non-deceptive | Holistic | Deepfakes are artificial face swaps or identity alterations used to impersonate individuals. | 4 |
| 361 | 2024 | Ekpang et al. | A→B | Deceptive | Targeted and Holistic | Deepfakes are artificial images or videos that manipulate identities, often used for deception. | 6 |
| 362 | 2023 | Falahkheirkhah et al. | A→B | Deceptive and Non-deceptive | Holistic | Deepfakes in this context refer to synthetic histologic images generated using deep learning techniques to augment digital pathology training and diagnostic tools. | 33 |
| 363 | 2023 | Fehring and Bonaci | A→B | Deceptive | Holistic | An image or signal generated by deep learning that swaps the face, speech, or actions of one subject with another. | 4 |
| 364 | 2023 | Firc et al. | A→B | Deceptive | Holistic | Synthetic media designed to deceive AI systems into accepting fake identities. | 42 |
| 365 | 2023 | Fuad et al. | A→B | Deceptive | Holistic | Deepfakes are synthetic media created using AI and machine learning to alter or swap faces in videos. | 2 |
| 366 | 2023 | Gadgilwar et al. | A→B | Deceptive | Targeted | Deepfakes are artificial manipulations of images or videos to impersonate individuals, often for deceptive purposes. | 3 |
| 367 | 2023 | Gao et al. | Mixed Types | Deceptive | Holistic | Deepfakes are artificial media content created using AI and machine learning techniques to replace or superimpose faces in existing images or videos with other people's faces. | 6 |
| 368 | 2023 | Geng | A→B | Deceptive | Targeted | Artificially generated content used for various purposes including entertainment, fraud, and manipulation. | 12 |
| 369 | 2023 | George and George | A→B | Deceptive | Targeted and Holistic | Synthetic media created using AI and machine learning techniques to produce highly realistic fake videos or audio recordings. | 16 |
| 370 | 2023 | Getman and Yilan | A→B | Non-deceptive | Holistic | Deepfakes are artificial AI-generated manipulations of data, such as images, videos, or audio, used to alter or fake information. | 3 |

| 371 | 2023 | Gil et al. | A→B | Deceptive | Holistic | Technology involving AI methods to create realistic fake content by manipulating identities or attributes. | 32 |
|---|---|---|---|---|---|---|---|
| 372 | 2023 | Giri and Brady | A→B | Deceptive | Holistic | Synthetic videos used to enhance communication or educational purposes, particularly tailored for people with Autism. | 11 |
| 373 | 2023 | Glick | A→B | Deceptive and Non-deceptive | Holistic | AI-generated satirical videos using realistic faces to depict individuals or events. | 12 |
| 374 | 2023 | Gosztonyi and Lendvai | A→B | Deceptive | Holistic | Deepfakes are technologies used to create misleading information through face swapping or attribute modification. | 7 |
| 375 | 2023 | Govindu et al. | A→B | Deceptive | Holistic | Artificially generated audio that imitates human voices for fraudulent purposes. | 3 |
| 376 | 2023 | Guefrachi et al. | A→B | Deceptive | Holistic | Artificially generated videos that use AI to create realistic fake content, often involving face swapping or identity replacement. | 11 |
| 377 | 2023 | Gupta et al. | A→B | Deceptive | Holistic | Artificially generated images or videos that mimic real-life scenarios for deceptive purposes. | 44 |
| 378 | 2023 | Ha et al. | A→B | Deceptive | Holistic | Artificially generated videos that involve face swapping or alterations to mislead viewers. | 7 |
| 379 | 2023 | Hailtik and Afifah | A→B | Deceptive | Holistic | AI-generated deepfakes using machine learning to alter or swap identities for fraudulent purposes. | 7 |
| 380 | 2023 | Han et al. | A→B | Deceptive and Non-deceptive | Holistic | Generated images using GANs for face manipulation to deceive or impersonate individuals. | 19 |
| 381 | 2023 | Heo et al. | A→B | Non-deceptive | Targeted and Holistic | Artificially generated videos that manipulate identities through face swapping or attribute modification to deceive. | 87 |
| 382 | 2023 | Hou et al. | A→B | Deceptive and Non-deceptive | Targeted | Artificially generated virtual faces that can be animated using AI models trained on existing data to replicate or modify facial features and expressions. | 5 |
| 383 | 2023 | Ikram et al. | A→B | Deceptive | Targeted | Deepfakes used for detection purposes involve AI models identifying manipulated videos, often in security contexts like fraud prevention or identity verification. | 29 |
| 384 | 2023 | Jiang et al. | A→B | Deceptive | Holistic | Deepfakes are artificially generated images that manipulate human faces using AI techniques for deceptive purposes. | 1 |
| 385 | 2023 | Kılıç and Kahraman | A→B | Deceptive | Targeted | Deepfakes are AI-generated manipulations that involve realistic identity replacement or attribute modification. | 9 |
| 386 | 2023 | Kaate et al. | A→B | Non-deceptive | Holistic | Deepfakes are artificial images or videos that manipulate identities through techniques like face swapping and attribute modification to create realistic personas. | 17 |
| 387 | 2023 | Kandari et al. | A→B | Deceptive | Holistic | Techniques used to create fake content, particularly involving multimedia data manipulation for deception. | 9 |
| 388 | 2023 | Karan and Angadi | A→B | Deceptive and Non-deceptive | Targeted | Deepfakes involve AI-generated images or videos that manipulate identities, such as face swapping. | 45 |
| 389 | 2023 | Kaswan et al. | A→B | Deceptive | Holistic | Artificially generated multimedia that realistically superimposes one person's likeness or voice onto another using AI techniques. | 6 |
| 390 | 2023 | Kawa et al. | A→B | Deceptive and Non-deceptive | Holistic | The paper discusses audio-based deepfakes that involve manipulating voice samples to deceive systems into believing the audio is genuine. | 47 |
| 391 | 2023 | Kerenalli et al. | A→B | Deceptive | Holistic | Deepfakes are images or videos where identities have been manipulated through techniques such as face swapping or attribute modification. | 3 |
| 392 | 2023 | Khan et al. | A→B | Deceptive | Holistic | Artificially generated videos used for deception or manipulation. | 8 |

| 393 | 2023 | Khanjani et al. | A→B | Deceptive and Non-deceptive | Targeted and Holistic | Artificially generated audio content that manipulates voice or identity to deceive. | 80 |
|---|---|---|---|---|---|---|---|
| 394 | 2023 | Khatri et al. | A→A' | Deceptive | Holistic | Deepfakes are synthetic digital media that replace one identity with another, often through AI models to create realistic fake videos, images, or audio. | 21 |
| 395 | 2023 | Khder et al. | A→B | Deceptive | Holistic | Artificially generated fake videos that closely resemble real ones. | 11 |
| 396 | 2023 | Kilinc and Kaledibi | A→B | Non-deceptive | Holistic | Artificially generated audio sequences that mimic real voices with realistic features like intonation and accents. | 8 |
| 397 | 2023 | Kingra et al. | A→B | Deceptive | Holistic | Deepfakes are AI-based alterations of digital videos that allow the depiction of another person's behavior without their presence. | 31 |
| 398 | 2023 | Kleinlogel et al. | A→B | Deceptive | Holistic | Artificially generated datasets or content that may involve synthetic data manipulation, such as face swapping or identity replacement. | 2 |
| 399 | 2023 | Kosarkar et al. | A→B | Deceptive | Targeted | Artificially generated images or videos that manipulate face information to replace or substitute identities. | 60 |
| 400 | 2023 | Krätzer et al. | A→B | Deceptive | Holistic | AI-generated media where identities or content are altered to deceive. | 3 |
| 401 | 2023 | Krause et al. | A→B | Deceptive and Non-deceptive | Targeted | Deepfakes are videos that use phonemes, mouth movements, and video features to detect fake content, particularly in political contexts. | 1 |
| 402 | 2023 | Kshetri et al. | A→B | Non-deceptive | Holistic | Deepfakes are AI-generated digital content that mimics real-world information with high realism. | 3 |
| 403 | 2023 | Kuck | A→B | Deceptive | Holistic | The paper discusses generative AI models that create highly realistic imagery, often with manipulated identities or attributes, which can be used for various purposes including identity theft and objectification. | 4 |
| 404 | 2023 | Le and Woo | A→B | Deceptive | Holistic | AI-generated videos that manipulate identities for deceptive purposes. | 32 |
| 405 | 2023 | Le et al. | A→B | Non-deceptive | Holistic | Artificially generated media that realistically impersonates real content to deceive. | 9 |
| 406 | 2023a | Leone | A→B | Deceptive | Targeted and Holistic | AI-generated fakes used to simulate intelligent behaviors, particularly focusing on deception and identity replacement. | 10 |
| 407 | 2023b | Leone | A→B | Deceptive | Holistic | Deepfakes are artificially generated images or videos that manipulate digital faces, often for deceptive purposes. | 20 |
| 408 | 2023 | Lewis et al. | A→B | Deceptive | Holistic | Artificially generated videos that can mislead people into thinking they are authentic or involve someone else. | 7 |
| 409 | 2023 | Li et al. | Mixed Types | Deceptive | Targeted | The paper discusses voice deepfakes that involve audio-based manipulations such as modifying specific features like age and expression to deceive or impersonate individuals. | 18 |
| 410 | 2023 | Li et al. | A→B | Deceptive | Holistic | Deepfakes are synthetic face videos used to verify authenticity by modifying facial features and edges. | 1 |
| 411 | 2023 | Liao et al. | A→B | Deceptive | Holistic | Deepfakes are face manipulation techniques used to create realistic videos that involve altering or replacing people's identities, often for malicious purposes such as fraud or identity theft. | 58 |
| 412 | 2023 | Liu et al. | A→B | Deceptive | Holistic | Deepfakes are artificial videos with manipulated audio and visuals to appear realistic, often involving face swapping or attribute modification. | 18 |
| 413 | 2023 | Lu and Chu | A→B | Non-deceptive | Holistic | Digital face swaps or realistic image manipulations used to create resurrection narratives. | 29 |
| 414 | 2023 | Lubowitz | A→B | Deceptive | Targeted | AI tools like ChatGPT that create new content tailored to user requests without summarizing existing information. | 125 |
| 415 | 2023 | Mahmud et al. | A→B | Deceptive | Targeted | Deepfakes are artificial videos where faces are swapped or altered to impersonate others. | 3 |

| 416 | 2023 | Mai et al. | A→B | Deceptive | Holistic | Deepfakes used for identity transfer, where attributes like faces are swapped or modified using deep learning models. | 69 |
|---|---|---|---|---|---|---|---|
| 417 | 2023 | Makki and Jawad | A→B | Non-deceptive | Holistic | AI-driven alterations to media content that may involve identity swapping or attribute changes. | 2 |
| 418 | 2023a | Mallet et al. | A→B | Deceptive | Holistic | Artificially generated images or videos that manipulate faces to impersonate others. | 11 |
| 419 | 2023b | Mallet et al. | A→B | Deceptive | Holistic | Synthetic faces used to mimic real identities for spreading false information. | 15 |
| 420 | 2023 | Mankoo | A→A' | Deceptive | Targeted | Artificially generated videos used to impersonate people or spread false information, often with realistic features. | 2 |
| 421 | 2023 | Mary and Edison | A→B | Deceptive | Targeted | AI-generated realistic videos or images altered to impersonate others. | 19 |
| 422 | 2023 | Masood et al. | Mixed Types | Deceptive and Non-deceptive | Targeted and Holistic | Deepfakes are fabricated or altered media content created using deep learning techniques for deceptive purposes. | 510 |
| 423 | 2023 | Masud et al. | A→B | Deceptive | Holistic | Artificially generated videos that use deep learning techniques to alter or disguise the original content. | 14 |
| 424 | 2023 | Mathews et al. | A→B | Deceptive | Holistic | Deepfakes are images that have been manipulated to include fake identities or altered content, often for fraudulent purposes. | 20 |
| 425 | 2023 | Matygov et al. | A→B | Non-deceptive | Holistic | Artificially generated content that realistically swaps or replaces identities between individuals. | 4 |
| 426 | 2023 | Mekkawi et al. | A→B | Deceptive | Holistic | Deepfakes are AI-generated fake identities used for deception and fraud. | 3 |
| 427 | 2023 | Mira | A→B | Deceptive | Holistic | Artificially generated videos that manipulate or replace parts of real content to deceive. | 10 |
| 428 | 2023 | Misirlis and Munawar | A→B | Deceptive | Targeted and Holistic | Deepfakes are artificial images or videos that replace parts of real media with synthetic ones, often featuring faces. | 7 |
| 429 | 2023 | Mubarak et al. | A→B | Deceptive | Holistic | Deepfakes are AI-generated manipulations that can be used to create fake visual, audio, or textual content, often involving face swapping or attribute modification to deceive users. | 58 |
| 430 | 2023 | Mukta et al. | A→B | Deceptive | Holistic | Technology that creates artificial videos or images with altered identities. | 40 |
| 431 | 2023 | Muppalla et al. | A→B | Deceptive | Holistic | Deepfakes are AI-generated media where images or videos are modified to create convincing fake content. They often involve altering a person's face in a video to impersonate another individual. | 32 |
| 432 | 2023 | Murillo-Ligorred et al. | A→B | Deceptive | Targeted and Holistic | Artificially generated images that mimic real appearances or expressions. | 19 |
| 433 | 2023 | Mustak et al. | A→B | Deceptive | Targeted and Holistic | Deepfakes are artificially generated images or videos that mimic real-life scenarios and individuals, often used for deception. | 158 |
| 434 | 2023 | Nailwal et al. | A→B | Deceptive | Holistic | Artificially generated images or videos that manipulate identities, such as face swapping, to deceive users. | 3 |
| 435 | 2023 | Naitali et al. | A→B | Non-deceptive | Holistic | Deepfakes are artificial digital content created using AI algorithms that mimic real-world scenarios, identities, or events with high realism. They can involve face swapping, attribute modification, or complete synthetic faces. | 71 |
| 436 | 2023 | Nautiyal et al. | A→B | Deceptive | Holistic | Not explicitly defined in the paper; AI's role is described without a clear definition of deepfakes. | 7 |
| 437 | 2023 | Nawaz et al. | A→B | Deceptive | Targeted | Artificially generated fake videos that swap or alter human identities to deceive. | 29 |
| 438 | 2023 | Nnamdi et al. | A→B | Deceptive | Targeted and Holistic | AI-generated fake videos/images used to deceive. | 8 |
| 439 | 2023 | Okolie | A→B | Deceptive and Non-deceptive | Holistic | Deepfakes are images or recordings that have been altered to appear realistic but involve the replacement of identities. | 51 |

| 440 | 2023 | AK | A→B | Deceptive | Targeted | Artificially generated videos that involve face swapping to change identities with realistic results. | 4 |
|---|---|---|---|---|---|---|---|
| 441 | 2023 | Painter | A→B | Deceptive and Non-deceptive | Holistic | Deepfakes are artificial-generated videos or audio that mimic real people's voices or expressions, often used for deceptive purposes. | 9 |
| 442 | 2023 | Palmiotto | A→B | Deceptive and Non-deceptive | Holistic | AI-generated content designed to deceive by altering or impersonating identities. | 5 |
| 443 | 2023 | Parikh et al. | A→B | Deceptive | Targeted | Artificially generated audiovisual content that manipulates and alters real-world data to spread false information. | 6 |
| 444 | 2023 | Parslow | A→B | Non-deceptive | Holistic | Deepfakes used for artistic exploration of diverse identities through AI technologies such as facial recognition software. | 6 |
| 445 | 2023 | Patel et al. | A→B | Deceptive and Non-deceptive | Holistic | Deepfakes are synthetic images, videos, or audio streams created using AI techniques like GANs to impersonate individuals or spread false information. | 95 |
| 446 | 2023 | Patel et al. | A→B | Deceptive | Targeted | Deepfakes are artificial images or videos where identities have been manipulated, often through face swapping or attribute changes. | 85 |
| 447 | 2023a | Patel et al. | A→B | Deceptive | Targeted and Holistic | AI-altered videos meant to disseminate false information about individuals. | 5 |
| 448 | 2023b | Patil et al. | A→A' | Deceptive | Targeted | Deepfakes are realistic images or videos used to deceive by altering biological features such as faces, eyes, heartbeats, and expressions. | 14 |
| 449 | 2023 | Powers et al. | A→B | Deceptive | Holistic | Synthetic video avatars used in marketing contexts to create realistic appearances of individuals. | 9 |
| 450 | 2023 | Pryor et al. | A→B | Deceptive | Holistic | Deepfakes are images or videos where one's face is replaced with another's to spread misinformation. | 11 |
| 451 | 2023 | Punyani and Chhikara | A→B | Deceptive and Non-deceptive | Targeted and Holistic | Deepfakes involve the creation of synthetic images or videos with realistic facial features, often used to deceive or impersonate individuals. | 1 |
| 452 | 2023 | Quadir et al. | A→B | Deceptive | Holistic | Deepfakes are artificial images or videos that use machine learning to superimpose one person's face over another. | 3 |
| 453 | 2023 | Rafique et al. | A→B | Deceptive | Targeted | Deep fakes are artificially generated images or videos that manipulate attributes such as age, expression, or identity to spread disinformation. | 94 |
| 454 | 2023 | Rajalaxmi et al. | A→B | Deceptive | Holistic | Deepfakes involve the creation of fake photos and movies where a face is altered or replaced to impersonate someone else. | 9 |
| 455 | 2023 | Saravana Ram et al. | A→B | Deceptive | Targeted | Artificially generated images or videos that use AI to create fake content, often for deceptive purposes. | 31 |
| 456 | 2023 | Ramachandran et al. | A→B | Deceptive | Holistic | Deepfakes are artificial images or videos that manipulate identities to deceive or alter perceptions. | 3 |
| 457 | 2023 | Rancourt-Raymond and Smaili | A→B | Deceptive | Targeted | Artificial face-swapping technology used to deceive individuals in fraudulent contexts. | 59 |
| 458 | 2023 | Rebello et al. | A→B | Deceptive | Holistic | Artificially generated videos used for fraudulent purposes like identity theft and reputation management. | 2 |
| 459 | 2023 | Ritter et al. | A→B | Deceptive | Holistic | Artificially generated fake videos or images that mimic real content for deceptive purposes. | 6 |
| 460 | 2023 | Roy and Raval | A→B | Deceptive | Holistic | AI-generated fake visual content that mimics real data to deceive or mislead users. | 4 |
| 461 | 2023 | Ruiz et al. | A→B | Deceptive | Holistic | Artificially generated images that manipulate identities for deceptive purposes. | 12 |
| 462 | 2023 | Sadiq et al. | A→B | Deceptive and Non-deceptive | Holistic | Machine-generated content used to deceive or manipulate public opinion on social platforms. | 25 |
| 463 | 2023 | Salman and Shamsi | A→B | Deceptive | Targeted | Artificially generated audio and video content used to deceive or spread harmful information. | 7 |

| 464 | 2023 | Samoilenko and Suvorova | A→A' | Deceptive | Targeted | Deepfakes are used to create realistic fake videos that manipulate public perception or identity. | 27 |
|---|---|---|---|---|---|---|---|
| 465 | 2023 | Shaaban et al. | A→B | Deceptive and Non-deceptive | Targeted | Artificial voice recordings that may manipulate attributes like speaking rate or tone to deceive. | 27 |
| 466 | 2023 | Shanthi et al. | A→B | Deceptive | Holistic | Deepfakes are synthetic images or videos that mimic real identities or scenarios, often used for deceptive purposes. | 13 |
| 467 | 2023 | Sharma | A→B | Deceptive | Targeted | AI-generated hyper-realistic videos used to impersonate people by showing them saying and doing things that never happened. | 2 |
| 468 | 2023a | Sharma et al. | A→B | Deceptive | Holistic | Artificially generated content that mimics real-world data with high realism, such as text from GPT and visuals from StyleGAN, often used for malicious purposes. | 15 |
| 469 | 2023b | Sharma et al. | A→B | Deceptive | Holistic | Artificially generated content created by replacing an original image or video with someone else's identity. | 33 |
| 470 | 2024 | Sharma et al. | A→B | Deceptive | Targeted and Holistic | Deepfakes are AI-generated videos that mimic real content, often used to deceive or impersonate individuals. | 4 |
| 471 | 2023 | Shoaib et al. | A→B | Deceptive | Holistic | Artificially generated content designed to deceive or mislead users by mimicking real information. | 86 |
| 472 | 2023 | Siegel et al. | A→B | Deceptive | Holistic | Deepfakes are manipulated media used to deceive by altering or replacing identities in images or videos. | 3 |
| 473 | 2023 | Silva | A→B | Deceptive | Targeted | Deepfakes are artificial constructions that create fake images or identities using AI and software to mimic real credentials or personas. | 3 |
| 474 | 2023 | Singh et al. | A→B | Deceptive | Targeted | Deepfakes are AI-generated techniques used to replace human actors in movie scenes by altering facial features and expressions. | 6 |
| 475 | 2023 | Sohan et al. | A→B | Deceptive and Non-deceptive | Targeted | Artificially generated videos that manipulate identities or attributes to deceive. | 6 |
| 476 | 2023 | Soleimani et al. | A→B | Deceptive | Holistic | Deepfakes are altered video contents generated by GANs that can be used for fictitious news, identity theft, political manipulation, and pornography. | 4 |
| 477 | 2023 | Somoray and Miller | A→B | Deceptive | Holistic | Artificially generated media manipulated to resemble real ones, possibly for deceptive purposes. | 30 |
| 478 | 2023 | Sontakke et al. | A→B | Deceptive | Holistic | AI-generated fake videos used for various malicious purposes such as spreading misinformation, impersonation, and creating fake news. | 8 |
| 479 | 2023 | Stavola and Choi | Mixed Types | Deceptive and Non-deceptive | Targeted and Holistic | Deepfakes are used to deceive users in virtual environments by altering digital media such as images or videos. | 11 |
| 480 | 2023 | Story and Jenkins | A→B | Deceptive | Holistic | Deepfakes are artificial images or videos that use deep learning to create non-veridical representations, particularly of faces, often for deceptive purposes. | 10 |
| 481 | 2023 | Stroebel et al. | A→A' | Deceptive | Targeted | Artificially generated material that involves face swapping or identity replacement using consumer-grade hardware. | 36 |
| 482 | 2023 | Tak | A→B | Deceptive | Holistic | Deepfakes are synthetic audio recordings that modify or alter voice characteristics to impersonate individuals. | 3 |
| 483 | 2023 | Tan et al. | A→B | Deceptive and Non-deceptive | Holistic | Artificially generated videos that are used to detect deepfakes, ensuring authenticity in video content. | 16 |
| 484 | 2023 | Tariq et al. | A→A' | Deceptive | Targeted | Artificially generated deepfakes that realistically swap faces between individuals to impersonate others. | 41 |
| 485 | 2023 | Thompson | A→B | Deceptive | Holistic | AI-generated fake videos that alter audio or visual attributes without full face replacement. | 7 |
| 486 | 2023 | Tiwari et al. | A→B | Deceptive | Holistic | AI-generated media that closely mimics real content to be difficult to distinguish without specialized tools. | 16 |

| 487 | 2023 | Tran et al. | A→B | Deceptive | Holistic | Deepfakes are realistic images or videos where identities have been manipulated. | 6 |
|---|---|---|---|---|---|---|---|
| 488 | 2023 | Tripathi et al. | A→B | Deceptive | Holistic | Deepfakes are technologies that create convincing synthetic content, such as images or videos, often used to deceive by mimicking real content. | 2 |
| 489 | 2023 | Truong et al. | A→B | Deceptive | Holistic | AI-generated realistic videos that alter identities or expressions through face swapping, lip-sync reenactment, and voice cloning. | 2 |
| 490 | 2023 | Twomey et al. | A→A' | Deceptive | Targeted | Artificially generated images or videos that swap faces between individuals to create false narratives or manipulate public perception. | 55 |
| 491 | 2023 | Ulutas et al. | A→B | Deceptive | Holistic | Artificially generated audio files that are manipulated to appear as if they were recorded by someone else. | 18 |
| 492 | 2023 | Vajpayee et al. | A→B | Deceptive | Targeted and Holistic | Deepfakes are manipulated images of human faces that appear realistic but are not genuine. | 5 |
| 493 | 2023 | Venkatachalam et al. | A→B | Deceptive | Targeted | Artificially generated images or videos that involve face swapping or identity impersonation for deception. | 5 |
| 494 | 2023 | Wang and Chow | A→B | Deceptive | Holistic | Deepfakes are synthetic videos that mimic real-life events or identities with traces of their creation left behind. | 79 |
| 495 | 2023 | Whittaker et al. | A→B | Deceptive | Holistic | Artificially generated videos that realistically swap faces between people or replace attributes to deceive. | 63 |
| 496 | 2023 | Wu et al. | A→B | Deceptive | Holistic | Deepfakes are artificial images or videos that use deep learning to create realistic synthetic content, often involving face swapping or attribute modification. | 17 |
| 497 | 2023 | Xu et al. | A→A' | Deceptive | Targeted | AI-generated media aiming to deceive by altering identities or attributes. | 31 |
| 498 | 2023 | Yi et al. | A→A' | Deceptive | Targeted | Artificially generated audio that manipulates voice characteristics to deceive or impersonate individuals. | 90 |
| 499 | 2023 | Yu et al. | A→B | Deceptive | Holistic | Deepfakes are videos where identities or attributes have been manipulated to deceive. | 36 |
| 500 | 2023 | Zhang et al. | A→B | Deceptive | Holistic | Deepfakes involve AI-generated manipulations of images to deceive or alter identities. | 11 |
| 501 | 2023 | Zhou et al. | A→B | Deceptive | Holistic | Artificially generated videos that are detected using frame-level methods with Bayesian inference weighting to identify deepfakes. | 3 |
| 502 | 2024 | Łabuz | A→B | Deceptive and Non-deceptive | Holistic | Artificially generated content that manipulates identities through deepfakes, often used for deceptive purposes. | 7 |
| 503 | 2024 | Łabuz and Nehring | A→B | Deceptive | Targeted and Holistic | Deepfakes are used to manipulate public opinion and spread disinformation during election campaigns. | 31 |
| 504 | 2024 | Ştefan et al. | Mixed Types | Deceptive | Targeted | Deepfakes are models that mimic artifacts from deepfake generators to deceive detection systems. | 1 |
| 505 | 2024 | Abbas and Taeihagh | A→B | Deceptive | Targeted | Deepfakes are artificially generated images or videos used to deceive individuals or societies. | 30 |
| 506 | 2024 | Abdullah et al. | A→B | Deceptive and Non-deceptive | Targeted and Holistic | Deepfakes are images produced using deep generative models that mimic real-world content, posing risks to online platforms. | 19 |
| 507 | 2024 | Agarwal and Ratha | A→B | Deceptive | Holistic | Artificially manipulated images where parts of the content are altered or swapped to deceive systems. | 12 |
| 508 | 2024 | Akhtar et al. | A→B | Deceptive | Targeted | Artificially synthesized video and audio content created using deep neural networks for various applications. | 14 |
| 509 | 2024 | Al-Adwan et al. | A→A' | Deceptive | Targeted | Artificially generated media designed to deceive individuals or systems by mimicking real content such as images, videos, or audio. | 14 |
| 510 | 2024 | Al-Qazzaz et al. | A→B | Deceptive | Holistic | Artificially generated face forgeries used for deception in digital media such as fraud and identity theft. | 5 |
| 511 | 2024 | Alanazi and Asif | A→B | Deceptive | Holistic | Deepfakes are videos, images, or audio that appear realistic and are created using AI algorithms. They often involve altering or swapping identities to create deceptive content. | 11 |

| 512 | 2024 | Alanazi et al. | A→B | Deceptive | Targeted | AI-generated images used for various applications including adult content and misinformation control. | 12 |
|---|---|---|---|---|---|---|---|
| 513 | 2024 | Alhaji et al. | A→B | Deceptive | Holistic | Artificially generated videos that use techniques like face swapping to deceive users. | 11 |
| 514 | 2024 | Alkishri et al. | A→B | Deceptive | Targeted | Deepfakes are images generated by GANs that mimic real-world faces or scenes, often used to deceive systems into believing they are authentic. | 14 |
| 515 | 2024 | Allen et al. | A→B | Deceptive and Non-deceptive | Targeted | Artificially generated videos designed to deceive viewers about their authenticity or content. | 12 |
| 516 | 2024 | Almestekawy et al. | A→B | Deceptive | Targeted | Artificially generated videos that manipulate digital content for deceptive purposes such as identity theft or fraud. | 6 |
| 517 | 2024 | Altuncu et al. | A→B | Deceptive | Holistic | AI-generated deepfakes that mimic or alter identities through technology, often used for deceptive purposes. | 31 |
| 518 | 2024 | Amerini et al. | A→B | Deceptive | Holistic | Deepfakes are synthetic media created using AI that mimic real content but are difficult to distinguish from genuine material. | 8 |
| 519 | 2024 | Arshed et al. | A→B | Deceptive | Targeted | AI-generated images with manipulated faces used for deception. | 29 |
| 520 | 2024 | Ba et al. | A→B | Deceptive | Holistic | High-fidelity fake videos that can be misleading or forged. | 41 |
| 521 | 2024 | Battista | A→A' | Deceptive and Non-deceptive | Targeted | AI-generated fake identities used for impersonation or influence. | 16 |
| 522 | 2024 | Bhagtani et al. | A→B | Deceptive | Targeted | Synthetic speech generated by AI models that can mimic real human voice for fraudulent purposes. | 7 |
| 523 | 2024 | Bokolo and Liu | Mixed Types | Non-deceptive | Targeted and Holistic | Artificially generated images or videos used to identify or impersonate individuals in social media forensics. | 12 |
| 524 | 2024 | Cai and Li | Mixed Types | Deceptive | Holistic | Deepfakes used to manipulate or splice audio clips within partially fake audio for forgery attacks. | 14 |
| 525 | 2024 | Cantero-Arjona and Sánchez-Macián | A→B | Deceptive | Holistic | Artificially generated content that realistically manipulates identities or features of individuals for deceptive purposes. | 2 |
| 526 | 2024 | Cardaş-Răduţă | A→B | Deceptive | Holistic | Not explicitly defined but inferred from context related to detecting fake news. | 5 |
| 527 | 2024 | Chakravarty and Dua | A→A' | Deceptive | Targeted | Not explicitly defined but likely involves audio manipulation for detection purposes. | 22 |
| 528 | 2024 | Chen | A→B | Deceptive | Holistic | Highly realistic fake content generated by AI technologies like machine learning (ML) and deep neural networks (DNNs). | 2 |
| 529 | 2024 | Chen et al. | A→B | Deceptive | Holistic | Artificially generated videos that involve face swapping or identity impersonation within compressed video data. | 6 |
| 530 | 2024 | Chitale et al. | A→B | Deceptive and Non-deceptive | Holistic | Artificially generated audio content that manipulates real-world sounds to deceive viewers or listeners. | 5 |
| 531 | 2024 | Choi et al. | A→B | Deceptive | Holistic | Artificially generated videos that use style latent vectors to mimic real faces or expressions for deceptive purposes. | 43 |
| 532 | 2024 | Deng et al. | A→B | Non-deceptive | Holistic | Artificially generated media that involves realistic alterations to identify forgeries created through deep learning techniques. | 19 |
| 533 | 2024 | Domenteanu et al. | A→B | Deceptive | Holistic | Deepfakes are artificial images or videos that manipulate identities through face swapping or attribute modification to deceive users. | 6 |
| 534 | 2024 | Edwards et al. | A→B | Non-deceptive | Holistic | Deepfakes are artificial techniques that generate images or videos to mimic real content, often used for deceptive purposes. | 6 |
| 535 | 2024 | Esoimeme | A→B | Non-deceptive | Targeted | AI-driven tools used for fraudulent activities such as money laundering and terrorism financing. | 4 |
| 536 | 2024a | Ferrara | A→B | Non-deceptive | Holistic | Deepfakes are images or videos used to spread disinformation and manipulate public opinion, particularly in elections. | 10 |
| 537 | 2024b | Ferrara | A→B | Non-deceptive | Holistic | Artificially generated content that impersonates real people or entities for deceptive purposes. | 203 |
| 538 | 2024 | Fitriana et al. | A→B | Deceptive and Non-deceptive | Holistic | Deepfakes are synthetic images where identities are manipulated and replaced using AI technology. | 1 |
| 539 | 2024 | Fletcher et al. | A→B | Deceptive | Holistic | Deepfakes involve creating convincing personas through technology, often used to deceive users. | 7 |
| 540 | 2024 | Gal and Bulgurcu | A→B | Deceptive and Non-deceptive | Holistic | Deepfakes are used to manipulate media images by altering or replacing identities through illustrative means. | 2 |
| 541 | 2024 | Gambín et al. | A→B | Deceptive | Holistic | Deepfakes involve the use of AI to manipulate images or videos to impersonate individuals. | 71 |
| 542 | 2024 | Gao et al. | A→B | Deceptive and Non-deceptive | Holistic | Deepfakes are artificial images or videos that use deep learning to manipulate textures and artifacts to create realistic fake identities or scenes. | 17 |
| 543 | 2024 | Garriga et al. | A→B | Deceptive | Holistic | Deepfakes are technologies used in disinformation strategies to deceive and manipulate public perception. | 8 |
| 544 | 2024 | Ghita et al. | A→B | Deceptive | Holistic | AI-generated images with manipulated identities or features. | 3 |
| 545 | 2024 | Gilbert and Gilbert | A→B | Deceptive | Targeted and Holistic | Deepfakes are AI-generated manipulated audio-visual content that can deceive individuals into believing false information. | 12 |
| 546 | 2024 | Gong and Li | Mixed Types | Deceptive | Targeted and Holistic | Artificially generated images or videos that are used to deceive by replacing or modifying identities. | 18 |
| 547 | 2024 | Gowda et al. | A→B | Deceptive | Targeted | Synthetic images created to look real but are actually fake. | 4 |
| 548 | 2024 | Guarnera et al. | Mixed Types | Non-deceptive | Targeted | AI-generated images or videos used to deceive. | 40 |
| 549 | 2024 | Hameleers | A→B | Deceptive and Non-deceptive | Targeted | AI-driven audiovisual fabrication used for deception and disinformation. | 24 |
| 550 | 2024 | Hapid | A→B | Deceptive | Holistic | Deepfakes are used for malicious purposes such as fraud, identity theft, and spreading false information. | 4 |
| 551 | 2024 | Haq et al. | A→B | Deceptive | Holistic | Deepfakes are AI-generated content used to manipulate public opinion by substituting identities or altering attributes. | 8 |
| 552 | 2024 | Hariprasad et al. | A→B | Deceptive and Non-deceptive | Targeted | AI-based techniques to detect deepfakes in videos by analyzing specific regions like the lip area. | 8 |
| 553 | 2024 | Hashmi et al. | Mixed Types | Deceptive | Holistic | Artificially generated audiovisual content designed to deceive or mislead viewers about its authenticity. | 6 |
| 554 | 2024 | Heidari et al. | A→B | Deceptive | Targeted | Artificial face manipulations detected using deep learning techniques for various purposes such as fraud prevention and identity theft detection. | 168 |
| 555 | 2024 | Hsu | A→B | Non-deceptive | Holistic | Deepfakes are images where attributes such as identity are altered or replaced to deceive users. | 10 |
| 556 | 2024 | Ishrak et al. | A→B | Deceptive | Holistic | Artificially generated videos that can be used to deceive or impersonate individuals. | 6 |
| 557 | 2024 | Javed et al. | A→B | Deceptive | Targeted | Artificially generated videos that involve face swapping or identity manipulation to deceive viewers. | 12 |
| 558 | 2024 | Jenkins and Roy | A→B | Deceptive | Holistic | Deepfakes are synthetic images generated using DCGANs to create photorealistic biometric samples for various applications. | 11 |
| 559 | 2024 | Josan | A→B | Deceptive | Holistic | Artificially generated audio that imitates real voices for various purposes such as entertainment or research. | 7 |
| 560 | 2024 | Joshi and Nivethitha | A→B | Deceptive | Holistic | Artificially generated images that are used to detect and combat deep fakes, which involve deceptive alterations or manipulations of digital content. | 7 |

| 561 | 2024 | Ju et al. | A→A' | Deceptive | Holistic | Deepfakes are realistic images or videos where one person's likeness is replaced by another using deep learning technologies. | 43 |
|---|---|---|---|---|---|---|---|
| 562 | 2024a | Kaate et al. | A→B | Deceptive | Holistic | Deepfakes are artificial avatars created using AI technology to represent users with specific emotional expressions. | 2 |
| 563 | 2024b | Kaate et al. | A→B | Deceptive | Holistic | Deepfakes are artificial personas created using AI technology to study user perceptions in design tasks. | 8 |
| 564 | 2024 | Kalaiarasu et al. | Mixed Types | Deceptive | Targeted and Holistic | Artificially generated technologies that manipulate identities or attributes to deceive. | 4 |
| 565 | 2024 | Kang et al. | A→B | Deceptive | Targeted | Artificially generated audio deepfakes that manipulate voice characteristics to appear realistic. | 7 |
| 566 | 2024 | Karaköse et al. | A→B | Deceptive | Holistic | Deepfakes are artificial videos or images that trickily replace parts of real content to deceive viewers. | 31 |
| 567 | 2024 | Kaur et al. | A→B | Deceptive | Holistic | AI-generated fake videos used to spread false information. | 50 |
| 568 | 2024 | Kaushal et al. | Mixed Types | Deceptive | Targeted | Artificially generated videos that look realistic but are manipulated to deceive or spread false information. | 15 |
| 569 | 2024 | Khalid et al. | A→B | Deceptive | Holistic | AI-generated tweets used to fabricate sentiments about deepfakes for malicious purposes. | 10 |
| 570 | 2024 | Khan et al. | A→B | Deceptive | Targeted | Deepfakes are deep learning-based technologies used for digital forensics to detect and investigate fake content on social media platforms. | 20 |
| 571 | 2024 | Kharvi | A→B | Deceptive | Holistic | AI-generated manipulated digital content like realistic videos or images crafted to deceive decision-making processes. | 11 |
| 572 | 2024 | Kulangareth et al. | A→B | Deceptive | Targeted and Holistic | An AI-based method using deep learning to detect synthetic voices by analyzing speech pause patterns. | 8 |
| 573 | 2024 | Kumar and Kundu | A→B | Deceptive | Targeted and Holistic | Artificially generated multimedia content designed to deceive by altering or impersonating identities. | 5 |
| 574 | 2024 | Kumar et al. | A→B | Deceptive | Holistic | Deepfakes are artificial images or videos created using techniques like face swapping and facial expression recombination to deceive or mislead. | 6 |
| 575 | 2024 | Lad | A→B | Deceptive | Holistic | Deepfakes are synthetic media created using adversarial techniques, particularly GANs, to deceive users by mimicking real content. | 1 |
| 576 | 2024 | Lai et al. | A→B | Deceptive | Holistic | Deepfakes are artificial face-forge detection models that create realistic fake identities or modify existing ones to deceive. | 14 |
| 577 | 2024 | Lai et al. | A→A' | Deceptive | Targeted | Deepfakes are images or videos manipulated using deep learning techniques to alter faces, identities, or backgrounds, often for deceptive purposes. | 19 |
| 578 | 2024 | Layton et al. | A→B | Deceptive | Targeted | Synthetic speech used for deception, often created through deep learning models to mimic real human speech. | 1 |
| 579 | 2024 | Lees | A→B | Deceptive | Holistic | Deepfakes are used to swap faces or voices in documentaries, creating artificial identities that disrupt authenticity. | 7 |
| 580 | 2024 | Li et al. | A→B | Deceptive | Targeted | Deepfakes in this context refer to AI-generated audio content that mimics real conversations or sounds. | 4 |
| 581 | 2024 | Li et al. | A→B | Deceptive and Non-deceptive | Holistic | Deepfakes discussed are synthetic faces generated using frequency domain techniques to closely resemble wild fake face distributions. | 8 |
| 582 | 2024 | Liu et al. | A→B | Deceptive | Targeted | Artificially generated images that manipulate facial features to create realistic synthetic faces or images. | 5 |
| 583 | 2024 | Lu and Yuan | A→B | Deceptive | Targeted and Holistic | Deepfakes are artificially generated misrepresentations, often involving face swapping or identity replacement to deceive viewers. | 6 |
| 584 | 2024 | Maheshwari and Paulchamy | A→B | Deceptive | Targeted and Holistic | AI-generated manipulated images or videos used for deception. | 11 |

| 585 | 2024 | Maheshwari et al. | A→B | Deceptive | Holistic | Artificial alterations in images aimed at detecting authenticity issues. | 28 |
|---|---|---|---|---|---|---|---|
| 586 | 2024 | Maheshwari et al. | A→B | Deceptive | Holistic | Artificially generated digital content that manipulates or impersonates real identities to deceive. | 50 |
| 587 | 2024 | Maiano et al. | ∅→B | Deceptive and Non-deceptive | Holistic | Artificially generated images used for detection purposes, often involving AI tools like Chat-GPT or Midjourney. | 7 |
| 588 | 2024 | Mania | A→B | Deceptive | Holistic | Deepfakes are used to create non-consensual intimate images, often for abusive purposes. | 56 |
| 589 | 2024 | Maniyal and Kumar | A→B | Deceptive | Targeted | Deepfakes are artificially generated multimedia content that uses deep learning to create realistic fake images, videos, or voices. | 7 |
| 590 | 2024 | Manoranjitham and Swaroop | A→B | Deceptive | Holistic | Deepfakes are artificial images designed to deceive by mimicking authentic content through advanced AI models. | 3 |
| 591 | 2024 | McCosker | A→B | Deceptive | Targeted | Deepfakes involve the creation of synthetic media through visual datasets and machine learning techniques, often involving face swapping or identity replacement. | 35 |
| 592 | 2024 | Moreira et al. | ∅→B | Deceptive | Holistic | Artificially generated content that creates synthetic identities or manipulates existing ones for various purposes. | 2 |
| 593 | 2024 | Moreno | A→B | Deceptive | Holistic | Deepfakes are synthetic images or videos used to create misleading or harmful content by altering identities or attributes. | 69 |
| 594 | 2024 | Mu et al. | A→B | Deceptive and Non-deceptive | Holistic | Deepfakes are AI-generated images that mimic real faces used for deceptive purposes like fraud. | 2 |
| 595 | 2024 | Munir et al. | A→B | Deceptive | Holistic | Artificially generated audio that manipulates voice identity through spoofing attacks like Tacotron and VITS TTS. | 3 |
| 596 | 2024 | Mutmainnah et al. | A→B | Deceptive | Holistic | Deepfakes are technology used to create fake news or images by manipulating faces. | 5 |
| 597 | 2024 | Napshin et al. | A→B | Non-deceptive | Holistic | Deepfakes are technologies used to create convincing digital representations that can manipulate identities or attributes for various purposes. | 2 |
| 598 | 2024 | Nawaz et al. | A→B | Deceptive and Non-deceptive | Holistic | Deepfakes are artificially generated audiovisual content designed to appear real but are actually fake, often created using advanced techniques like Generative Adversarial Networks for deceptive purposes. | 18 |
| 599 | 2024 | Omar et al. | A→B | Deceptive | Holistic | Artificially generated videos that manipulate faces to impersonate individuals for malicious purposes such as identity theft or deception. | 19 |
| 600 | 2024 | Overton | A→B | Deceptive | Holistic | Artificially generated synthetic video and audio used to deceive or manipulate public perception. | 9 |
| 601 | 2024 | Passos et al. | A→B | Deceptive | Holistic | Content manipulated to impersonate someone else or mislead for deceptive purposes. | 87 |
| 602 | 2024 | Pellicer et al. | A→B | Deceptive | Targeted and Holistic | AI-generated images that mimic real data, often used to replace or impersonate identities. | 20 |
| 603 | 2024 | Putra et al. | A→B | Deceptive | Holistic | AI-generated videos that mimic real content for deceptive purposes. | 4 |
| 604 | 2024 | Qadir et al. | A→A' | Deceptive | Targeted | Artificially generated videos that involve face manipulation, lip synchronization, and other synthetic changes to deceive. | 19 |
| 605 | 2024 | Raman et al. | A→B | Deceptive | Holistic | Deepfakes used for detecting fake news involve AI-generated content designed to mimic real information, particularly through face swapping or identity alteration techniques. | 47 |
| 606 | 2024 | Ramluckan | A→B | Deceptive | Holistic | Artificially generated media that alters or replaces identities to deceive or spread harmful information. | 11 |

| 607 | 2024 | Ramon et al. | A→B | Deceptive | Targeted | Deepfakes are realistic images or videos used to impersonate individuals for fraudulent purposes. | 7 |
|---|---|---|---|---|---|---|---|
| 608 | 2024 | Rana et al. | Mixed Types | Deceptive and Non-deceptive | Targeted | AI-generated videos that blend one person's facial expressions onto another to create nearly indistinguishable content from reality. | 6 |
| 609 | 2024 | Renier et al. | Mixed Types | Deceptive | Targeted | AI-generated faces or expressions used to mimic real behavior in videos. | 11 |
| 610 | 2024 | Saha et al. | A→B | Deceptive | Holistic | Artificially generated audio content that mimics real speech or voices to impersonate individuals. | 9 |
| 611 | 2024 | Saini et al. | A→B | Deceptive | Holistic | Deepfakes are artificial images or videos created using deep learning and GANs to manipulate identities, often for fraudulent purposes. | 3 |
| 612 | 2024 | Salini and HariKiran | A→B | Deceptive | Holistic | Artificially generated videos used to deceive or mislead, often involving realistic face swapping between individuals. | 6 |
| 613 | 2024a | Samuel-Okon et al. | A→B | Deceptive | Holistic | Deepfakes are artificial face swaps or identity alterations used deceptively for various purposes. | 30 |
| 614 | 2024b | Samuel-Okon et al. | A→B | Deceptive | Holistic | AI-generated content used to deceive and spread false information, often mimicking real identities. | 50 |
| 615 | 2024 | Sandotra and Arora | A→B | Non-deceptive | Holistic | Artificially generated media aiming to deceive or mimic real content. | 17 |
| 616 | 2024 | Sayyed | A→B | Non-deceptive | Holistic | AI-generated content used for deceptive purposes, potentially involving identity manipulation in legal contexts. | 6 |
| 617 | 2024 | Schmitt and Flechais | A→B | Deceptive | Holistic | Generative AI-generated content used for malicious purposes such as identity impersonation and fraud. | 67 |
| 618 | 2024 | Seng et al. | A→B | Deceptive | Holistic | Deepfakes are technologies that modify audio and visual content to deceive users, often used in entertainment and social media contexts. | 9 |
| 619 | 2024 | Shahin and Deriche | A→B | Deceptive | Holistic | Synthetic images created using advanced AI techniques that are difficult to distinguish from real ones and often used for deception. | 2 |
| 620 | 2024 | Sharma et al. | A→B | Deceptive | Holistic | Artificially generated digital content that realistically alters or swaps identities in visual media. | 8 |
| 621 | 2024 | Shree et al. | A→B | Deceptive and Non-deceptive | Targeted | Artificially generated media that realistically swap or alter faces between people to deceive or impersonate others. | 3 |
| 622 | 2024 | Siegel et al. | A→B | Deceptive | Holistic | Deepfakes are AI-generated content used to detect or analyze other AI systems in media forensic contexts. | 10 |
| 623 | 2024 | Singer | A→B | Deceptive | Holistic | Artificially generated images that swap or modify the identities of individuals, often for harmful purposes. | 10 |
| 624 | 2024 | Songja et al. | A→B | Deceptive | Holistic | AI-generated images that can be manipulated or appear realistic enough to deceive. | 6 |
| 625 | 2024 | Soudy et al. | A→B | Deceptive | Holistic | Artificially generated videos that use techniques like face swapping or replacement to mimic real content. | 15 |
| 626 | 2024 | Sudarsan et al. | A→B | Deceptive | Holistic | Artificially generated content that realistically manipulates media to deceive or mislead. | 4 |
| 627 | 2024 | Sufian | A→B | Non-deceptive | Targeted | Artificially generated videos that use deep learning algorithms to create realistic content, often for deceptive purposes. | 3 |
| 628 | 2024 | Takale et al. | A→B | Deceptive | Holistic | AI-generated alterations used for creative and design purposes such as face swapping or attribute modification in images. | 25 |
| 629 | 2024 | Tan et al. | A→B | Deceptive | Holistic | Deepfakes are synthetic images or videos used for deceptive purposes, often involving altered identities. | 67 |
| 630 | 2024 | Tariq et al. | A→B | Deceptive | Holistic | Artificially generated videos that may involve face swapping or other manipulations to deceive users. | 3 |
| 631 | 2024 | Todupunuri | A→B | Deceptive | Holistic | Artificially generated images or videos that mimic real content for deceptive purposes. | 3 |

| 632 | 2024 | Tsigos et al. | A→B | Deceptive | Holistic | Deepfakes are artificial images or videos that use deep learning to manipulate or replace parts of real images or videos to deceive viewers. | 9 |
|---|---|---|---|---|---|---|---|
| 633 | 2024 | Uppal et al. | A→B | Deceptive | Holistic | Deepfakes are artificial face swaps or identity impersonations used for various purposes. | 3 |
| 634 | 2024 | Valente et al. | A→A' | Deceptive | Targeted | Artificially generated speech signals that are meant to mimic real audio voices for various purposes such as entertainment or education. | 4 |
| 635 | 2024 | Vashishtha et al. | A→B | Deceptive | Holistic | Artificially generated images and videos that are manipulated to deceive or impersonate individuals. | 3 |
| 636 | 2024 | Vig | A→B | Deceptive | Holistic | Artificially generated images or videos that involve realistic face swapping or identity replacement between individuals. | 2 |
| 637 | 2024 | VP and Dheepthi | A→B | Deceptive and Non-deceptive | Targeted | Deepfakes are synthetic media designed to deceive by altering or replacing identities. | 5 |
| 638 | 2024 | Wang and Huang | A→B | Deceptive and Non-deceptive | Holistic | Artificially generated audio-visual content that manipulates real-world features like faces and expressions to create realistic fake videos. | 6 |
| 639 | 2024 | Wang et al. | A→B | Deceptive | Holistic | Deepfakes created using spatial and frequency domain features to mimic real images or videos for forgery purposes. | 11 |
| 640 | 2024 | Warren et al. | Mixed Types | Deceptive | Targeted and Holistic | Artificially generated audio samples designed to deceive users about their origin or identity. | 8 |
| 641 | 2024 | Wazid et al. | A→B | Deceptive | Both Targeted and Holistic | Synthetic media generated through AI techniques to mimic real content for deceptive purposes. | 18 |
| 642 | 2024 | Wu et al. | A→B | Deceptive | Holistic | DeepFakes are artificial faces that mimic real people's identities through realistic DeepFakes. | 7 |
| 643 | 2024 | Xia et al. | A→B | Deceptive | Holistic | Deepfakes are synthetic images or videos created using deep learning to deceive by altering identities or features. | 5 |
| 644 | 2024 | Xie et al. | ∅→B | Deceptive | Holistic | Artificially generated audio aiming to deceive or replace original content. | 6 |
| 645 | 2024 | Xu et al. | Mixed Types | Deceptive | Holistic | Not explicitly defined but inferred to involve the use of gradient and content features for detection purposes. | 7 |
| 646 | 2024 | Yadav and Vishwakarma | A→B | Deceptive | Holistic | Artificially generated images or videos with manipulated faces used to deceive. | 13 |
| 647 | 2024 | Yang et al. | A→B | Deceptive | Holistic | Artificially generated images that swap or replace faces between individuals. | 4 |
| 648 | 2024 | Yang et al. | A→B | Deceptive | Holistic | Realistic counterfeit human facial images used for deceptive purposes such as fraud, identity theft, and explicit content distribution. | 9 |
| 649 | 2024 | Yu et al. | A→B | Deceptive | Targeted | Artificial audio manipulations aiming to impersonate real speech for deceptive purposes. | 12 |
| 650 | 2024 | Zhalgasbayev et al. | A→B | Deceptive | Holistic | Deepfakes are artificial images created using advanced AI models to mimic real faces or scenes, often for deceptive purposes. | 5 |
| 651 | 2024 | Zhang | A→B | Deceptive | Holistic | Deepfakes involve creating realistic fake videos using techniques like face swapping to replace or alter identities. | 1 |
| 652 | 2025 | Ahmad et al. | A→A' | Deceptive | Targeted | No clear definition provided as the paper does not discuss deepfakes specifically. | 24 |
| 653 | 2025 | Jacobsen | A→B | Non-deceptive | Holistic | Artificially generated content that involves face swapping or identity replacement, often used to create anxiety regarding their harmful effects. | 7 |
| 654 | 2025 | Ke et al. | A→B | Deceptive | Holistic | Deepfakes used for detecting fraudulent activities in online transactions by manipulating payment images to deceive systems. | 32 |
| 655 | 2025 | Lin et al. | A→B | Deceptive | Holistic | Artificially generated audio that manipulates voice characteristics or other attributes to deceive. | 12 |

| 656 | 2025 | Lin et al. | A→B | Deceptive | Holistic | Deepfakes are artificially generated images or videos that manipulate visual data to deceive users, often by altering or replacing faces. | 5 |
| 657 | 2025 | Rosca et al. | A→B | Deceptive | Targeted | AI-generated text that mimics human writing styles to identify specific authors or identities. | 5 |
| 658 | 2025 | Tan et al. | A→B | Deceptive and Non-deceptive | Holistic | Deepfakes are artificial images or videos that manipulate identities through techniques like face swapping to deceive. | 15 |
| 659 | 2025 | Weikmann et al. | A→B | Deceptive | Targeted | Technologies that manipulate audio-visual content to deceive. | 18 |

f"""You are an expert in deepfake research and academic paper analysis. Please analyze the following academic paper and extract specific information.

**Paper Information:**
- Title: {content['title']}
- Author: {content['author']}
- Filename: {content['filename']}
**Paper Content (Key Sections):**
{content['first_pages'][:2000]}

**Task:**
Please provide a JSON response with the following fields:
1. **title**: The actual title of the paper (if not found in metadata, extract from text)
2. **author**: The actual author(s) (if not found in metadata, extract from text)
3. **application**: What are the PRIMARY APPLICATIONS/USES of deepfakes that this paper discusses or mentions? This refers to what deepfakes are USED FOR, not what the paper studies. Examples might include:
    - Entertainment (movies, social media, fun apps)
    - Identity theft/fraud (malicious impersonation)
    - Political manipulation (fake political content)
    - Pornography (non-consensual intimate images)
    - Art/creativity (artistic expression)
    - Education (historical figure recreation)
    - Or any other specific application the paper mentions
    Write the actual application domain(s) as discussed in the paper, not predefined categories.
4. **definition of deepfake**:
    - If the paper explicitly defines deepfakes: Summarize their definition in 1-2 clear sentences
    - If no explicit definition but implicit understanding: Infer and summarize what the paper considers deepfakes to be in 1-2 sentences
    - If deepfakes are barely mentioned or no clear concept: Write "No clear definition provided"
    - DO NOT quote long passages, summarize concisely
5. **identity_source**: Based on the TYPE OF DEEPFAKES that this paper discusses or mentions (NOT what the paper's research method is), classify as:
    - "Original-to-Target Transfer": The deepfakes discussed involve face swapping, identity replacement (A→B)
    - "Original Modification": The deepfakes discussed involve attribute modification, expression change (A→A')
    - "Complete Synthesis": The deepfakes discussed involve generating completely artificial/non-existent faces (∅→B)
    - "Mixed Types": Paper discusses multiple types of deepfake identity manipulation
6. **intent**: Based on the PURPOSE/APPLICATIONS of the deepfakes that this paper discusses (NOT the paper's research intent), classify as:
    - "Deceptive": The deepfakes discussed are primarily for malicious, fraudulent, harmful purposes
    - "Non-deceptive": The deepfakes discussed are primarily for legitimate, entertainment, research purposes
    - "Both Deceptive and Non-deceptive": The paper discusses deepfakes used for both malicious and legitimate purposes
7. **manipulation granularity**: Based on the LEVEL OF MANIPULATION of the deepfakes discussed in the paper, classify as:
    - "Targeted Attribute": The deepfakes discussed modify specific features/attributes (age, expression, etc.)

**Figure A1. The Complete Prompt for PDF Analysis of Deepfake Papers.**